\documentclass[11pt]{amsart}

\usepackage{amsmath,amsthm,amssymb}
\usepackage{esint}             
\usepackage{bm}                
\usepackage{venndiagram}

\usepackage{a4wide}
\usepackage{geometry}
\usepackage{fancyhdr}
\usepackage[breaklinks]{hyperref}
\usepackage{url}
\usepackage[numbers,square,sort&compress]{natbib}

\graphicspath{{./}{figs/}}
\usepackage{subcaption}
\usepackage[all]{xypic}
\usepackage{tikz}
\usetikzlibrary{arrows}
\usetikzlibrary{decorations.markings}
\usetikzlibrary{decorations.pathreplacing}
\usetikzlibrary{patterns}
\usetikzlibrary{shapes.geometric}
\usepackage{tikz-3dplot}

\usepackage{alltt}             
\usepackage{enumerate}
\usepackage{multicol}
\usepackage{comment}
\usepackage{arydshln}

\newcounter{comments}
\usepackage{color}
\definecolor{darkblue}{rgb}{0, 0, .6}
\definecolor{grey}{rgb}{.7, .7, .7}
\hypersetup{colorlinks=true,linkcolor=darkblue,anchorcolor=darkblue,
  citecolor=darkblue,pagecolor=darkblue,urlcolor=darkblue,
  pdftitle={},pdfauthor={}
}

\theoremstyle{definition}

\def\<{\langle}
\def\>{\rangle}

\def\And{\wedge}
\def\Or{\vee}
\newcommand{\Not}[1]{\overline #1}

\def\longto{\longrightarrow}

\def\F{\mathbb{F}}

\colorlet{lightred}{red!30!white}
\colorlet{lightblue}{blue!30!white}
\colorlet{lightyellow}{yellow!30!white}
\colorlet{keylime}{green!10!white}
\colorlet{darkgreen}{green!50!black}

\tikzset{
  activ/.style={
    decoration={markings,mark=at position 1 with {\arrow[scale=2]{stealth}
      }
    }, 
    postaction={decorate}
  }
}
\tikzset{
    inhib/.style={
      shorten >=#1,
      decoration={
        markings,
        mark={
          at position 1
          with {
            \draw[fill] circle [radius=#1];
          }
        }
      },
      postaction=decorate
    },
    inhib/.default=1.75pt
}

\begin{document}

\author{Isadora Deal}
\address{School of Medicine, University of South Carolina, Columbia, SC, 29209}
\email{Isadora.Deal@uscmed.sc.edu}

\author{Matthew Macauley}
\address{School of Mathematical and Statistical Sciences, Clemson University, Clemson, SC 29634} 
\email{macaule@clemson.edu}

\author{Robin Davies}
\address{Radford University Carilion, Roanoke, VA 24013}
\email{rdavies@radford.edu}

\title[Boolean models of the \emph{trp} and \emph{tna} operons]{Boolean models of the transport, synthesis, and metabolism of tryptophan in \emph{Escherichia Coli}}


\maketitle

\begin{abstract}
    The tryptophan (\emph{trp}) operon in \emph{E. coli} codes for the proteins responsible for the synthesis of the amino acid tryptophan from chorismic acid, and has been one of the most well-studied gene networks since its discovery in the 1960s. The tryptophanase (\emph{tna}) operon codes for proteins needed to transport and metabolize it. Both of these have been modeled individually with delay differential equations under the assumption of mass-action kinetics. Recent work has provided strong evidence for bistable behavior of the \emph{tna} operon. The authors of \cite{orozco2019bistable} identified a medium range of tryptophan in which the system has two stable steady-states, and they reproduced these experimentally. In this paper, we will show how a Boolean model can capture this bistability. We will also develop and analyze a Boolean model of the \emph{trp} operon. Finally, we will combine these two to create a single Boolean model of the transport, synthesis, and metabolism of tryptophan. In this amalgamated model, the bistability disappears, presumably reflecting the ability of the \emph{trp} operon to produce tryptophan and drive the system toward homeostasis. All of these models have longer attractors that we call ``artifacts of synchrony'', which disappear in the asynchronous automata. This curiously matches the behavior of a recent Boolean model of the arabinose operon in \emph{E. coli}, and we discuss some open-ended questions that arise along these lines.
\end{abstract}

\section{Introduction}

An \emph{operon} is a cluster of genes that collectively serve a common  purpose and are transcribed together. We say than an operon is ``on'' if its genes are being transcribed at a high \emph{expression level} (rate), and ``off'' if transcription is being repressed, which can be achieved through a variety of mechanisms -- by a repressor protein or a DNA loop that physically blocks the RNA polymerase from transcribing the genes, or through the absence of or conformational change to a necessary transcription factor. Operons are a basic form of gene regulation, and are primarily used in prokaryotes. The lactose (\emph{lac}) operon in \emph{Escherichia coli} (\emph{E. coli}) was the first operon discovered \cite{jacob1960l'operon}. It regulates the genes that control the transport and metabolism of the sugar lactose in \emph{E. coli}, a bacterium and model organism, which lives in the gut of mammals and birds. This is an example of an \emph{inducible} operon, because it is normally off, and is turned on only when needed -- namely, when lactose is available but glucose, the prefered carbon source, is not. Moreover, the genes in the \emph{lac} operon are negatively controlled by a repressor protein. The arabinose (\emph{ara}) operon in \emph{E. coli} controls the transport and metabolism of the sugar arabinose, which the cell can also use as a carbon source, if glucose is not present. This is also an inducible operon, but the genes are positively controlled. The opposite of an inducible operon is a \emph{repressible} one, which is normally on unless the gene products are no longer needed. The first repressible operon discovered was the tryptophan (\emph{trp}) operon in \emph{E. coli}, in the early 1960s \cite{matsushiro1965transcription}. Tryptophan is one of the 20 amino acids that make up the building blocks of proteins. Organisms that feed on others get sufficient tryptophan from their diets, and as such, have not retained the ability to synthesize it. In contrast, organisms such as \emph{E. coli} need the ability to synthesize tryptophan, and \emph{E. coli} does this with the proteins coded by the structural genes in the \emph{trp} operon. However, if tryptophan is readily available, then it would be a waste of valuable cellular energy to transcribe those genes. Therefore, this operon is on by default, and is only turned off if it is not needed. The operon is turned off by a repressor protein that blocks transcription by attaching to the operator region of the operon. However, this protein needs to bind to tryptophan before it can bind to the operon. In this role, tryptophan is said to be a \emph{co-repressor}. The \emph{trp} operon is an example of a repressible operon under negative control. The fourth possibility, of a repressible operon under positive control, has never been found, and may not exist. The following table summarizes these four possibilities. 

\begin{table}[!ht]
\begin{tabular}{|l||c|c|} \hline
& inducible & repressible \\ \hline\hline
negative control & \emph{lac} & \emph{trp} \\ \hline
positive control & \emph{ara} & may not exist? \\ \hline
\end{tabular}
\caption{Examples of the four possible types of operons.}\label{tbl:operons}
\end{table}

Though many more operons have been discovered, the three listed in Table~\ref{tbl:operons} are the classic ones that are most widely known. All of these have been modeled with delay differential equations (DDE) \cite{mackey2004modeling,santillan2001dynamic,yildirim2004dynamics,yildirim2012mathematical}, and the \emph{lac} \cite{veliz-cuba2011boolean} and \emph{ara} operons \cite{jenkins2017bistability} have additionally been modeled with Boolean functions. In this paper, we will model the \emph{trp} operon with Boolean functions. We will do the same for the operon responsible for the transport and metabolism of tryptophan -- the tryptophanase (\emph{tna}) operon. A recent ODE model of that operon supports the hypothesis that it can exhibit bistability, i.e., the presence of multiple steady states (corresponding to the operon being on or off), under medium levels of tryptophan concentration. Our Boolean model will both provide more evidence to support this claim, as well as showing another example of how a complex but fundamental biological phenomenon such as bistability can be captured by a qualitative coarse-grained Boolean model. 

Though the \emph{trp} and \emph{tna} operons have previously been modeled separately, they are not independent entities. The latter codes for the TnaB protein, a membrane transport protein called a \emph{permease}, that transports tryptophan into the cell. This can have a direct impact on whether or not the \emph{trp} operon is needed to synthesize it. Moreover, the \emph{trp} repressor protein controls the Mtr protein, another permease that can transport tryptophan into the cell. A third permease, AroP, transports tryptophan, tyrosine, and phenylalanine. The three tryptophan permeases used by \emph{E. coli} are thought to not only transport tryptophan, but also maintain a high intracellular concentration when it is not available externally. However, the TnaB permease is the main transporter of tryptophan \cite{gu2013knocking}. Details of these three tryptophan permeases can be found in \cite{yanofsky1991physiological}. Due to all of this, a complete model of the symthesis, transport, and metabolism of tryptophan in \emph{E. coli} should incorporate features from both the \emph{trp} and \emph{tna} operons. We will propose such a model, by putting together our Boolean models of the individual operons. We will analyze our models using the BoolNet package in the R statistical programming language. For every choice of the parameters, we end up getting a unique fixed point that that we will explain biologically, except for the case where expect bistability, which has two fixed points. However, a number of these cases have other basins of attraction with longer limit cycles. Fortunately, these disappear if we allow the individual functions to be updated asynchronously, leaving only the fixed point(s) as attractors. We call this curious behavior an \emph{artifact of synchrony}, and it is something that also arose in a Boolean model of \emph{ara} operon, which exhibited bistability as well \cite{jenkins2017bistability}. In principle, there is no reason to expect this behavior, and we will conclude this paper with a discussion about open-ended questions on how and why this arises, and how to detect it using computational algebra -- a framework which has been used for a number of algorithms involving similar Boolean or algebraic models.

\section{The \emph{trp} and \emph{tna} operons}

Deoxyribonucleic acid (DNA) is composed of two antiparallel strands of nucleotides coiling around each other, and it carries all of the genetic information of an organism. The celebrated central dogma of molecular biology is the two-step process of \emph{transcription}, where the RNA polymerase enzyme synthesizes a strand of mRNA from a template strand of DNA, and  \emph{translation}, in which this mRNA is read by organelles called ribosomes, producing proteins, which are long chains of amino acids. Both eukaryotic and prokaryotic cells rely on proteins to carry out necessary functions such as growth and maintaining proper nutrient levels. 

Proteins are built from the twenty amino acids, which are molecules composed of three subunits connected to the same carbon: an amino group, a carboxylic acid group, and specific R groups. Amino acids are commonly abbreviated with a 3-letter code, or a shorter single letter code; for tryptohphan these are Trp and W. Tryptophan is found in most proteins and functions as the precursor of serotonin, melatonin, and vitamin B3. Its indole side chain makes it one of the ten amino acids that are non-polar.  

The \emph{trp} operon, which is responsible for synthesizing tryptophan when it is not available, consists of five structural genes. These code for the proteins and enzymes that make up the  biosynthesis pathway that converts chorismate into tryptophan. The \emph{tna} operon, which metabolizes tryptophan, consists of two structural genes. One of these codes for the tryptophanase enzyme involved in the tryptophan metabolic pathway, and the other codes for a permease, which transports tryptophan into the cell. 

\begin{figure}[!ht]
\begin{center}
    \includegraphics[width=.8\textwidth]{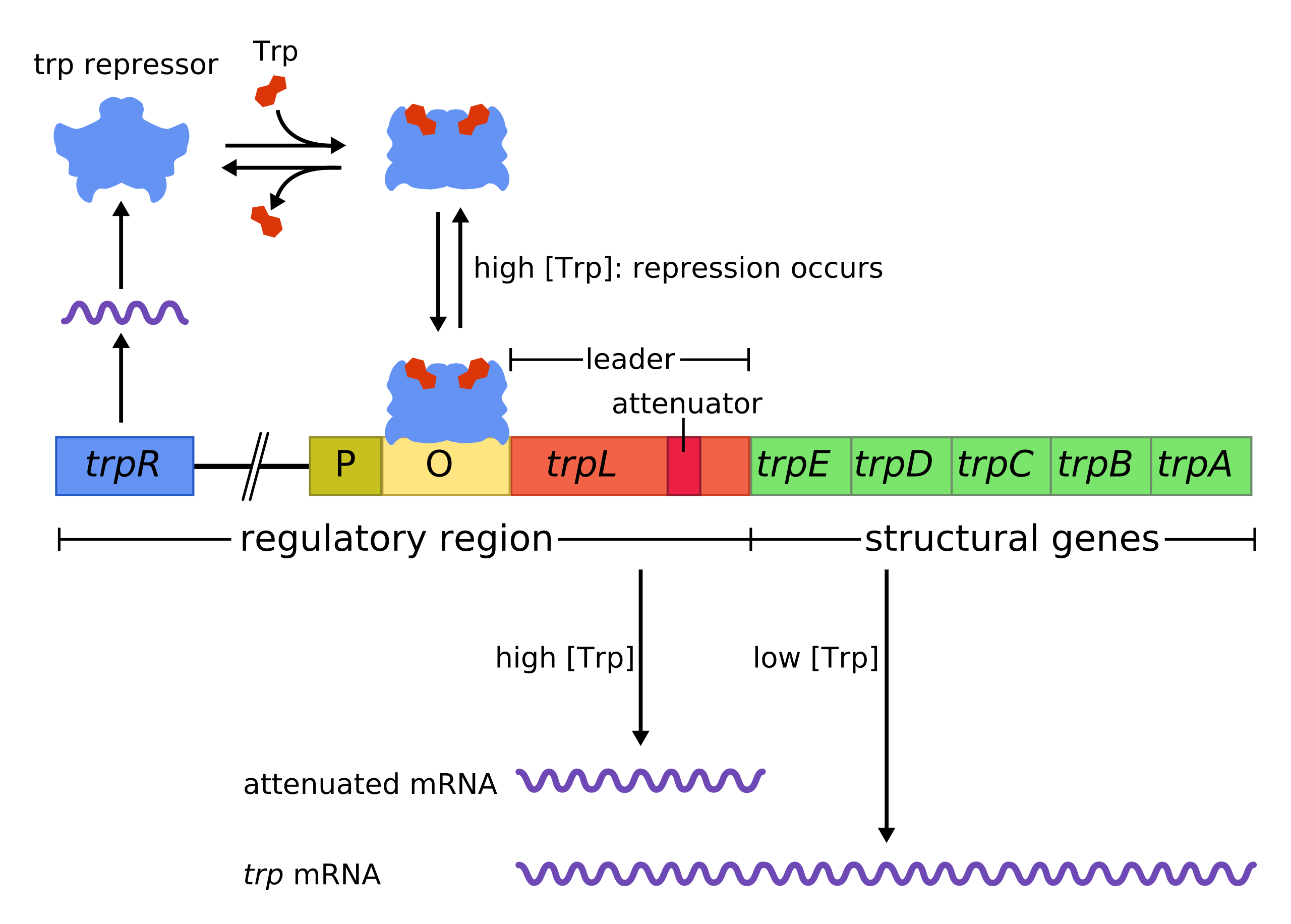}
\end{center}
\caption{The \emph{trp} operon in \emph{E.~coli}. P represents the promoter and O represents the operator. Image from Wikipedia, CC BY-SA 3.0.}\label{fig:trp-operon}
\end{figure}

Upstream of the structural genes in the \emph{trp} operon is the \emph{operator region}, where the repressor protein binds to block transcription. Immediately upstream of this is the \emph{promoter region}, where RNA polymerase binds to initiate transcription. Most operons have these three basic regions, though some more complex ones have more than one of these. The \emph{tna} operon has structural genes and a promoter, but not an operator region, because its regulation is controlled by different mechanisms. Figure~\ref{fig:trp-operon} shows the basic regions of the \emph{trp} operon. In the remainder of this section, we will give a biological overview of the \emph{trp} and the \emph{tna} operons. 

\subsection{Tryptophan synthesis and the \emph{trp} operon}

The tryptophan  (\emph{trp}) operon in \emph{E.~coli} is used to synthesize tryptophan when none is readily available. Its five structural genes, trpE, trpD, trpC, trpB, and trpA, code for the enzymes in the pathway that converts chorismic acid into tryptophan. The \emph{trp} operon utilizes three repression mechanisms to regulate gene expression levels. The most basic type is carried out by the TrpR repressor protein, which is coded for by the trpR region, upstream of the operon. Specifically, when \emph{E.~coli} cells contain high levels of tryptophan, it acts as a co-repressor by binding to the TrpR repressor protein. This binding causes a structural change, allowing the newly-formed protein complex to bind to the operator region, thereby preventing the RNA polymerase from moving downstream and transcribing the structural genes. This type of regulation can suppress transcription levels by a factor of 70 \cite{nelson2005principles}.

\begin{figure}[!ht]
\includegraphics[width=.515\textwidth]{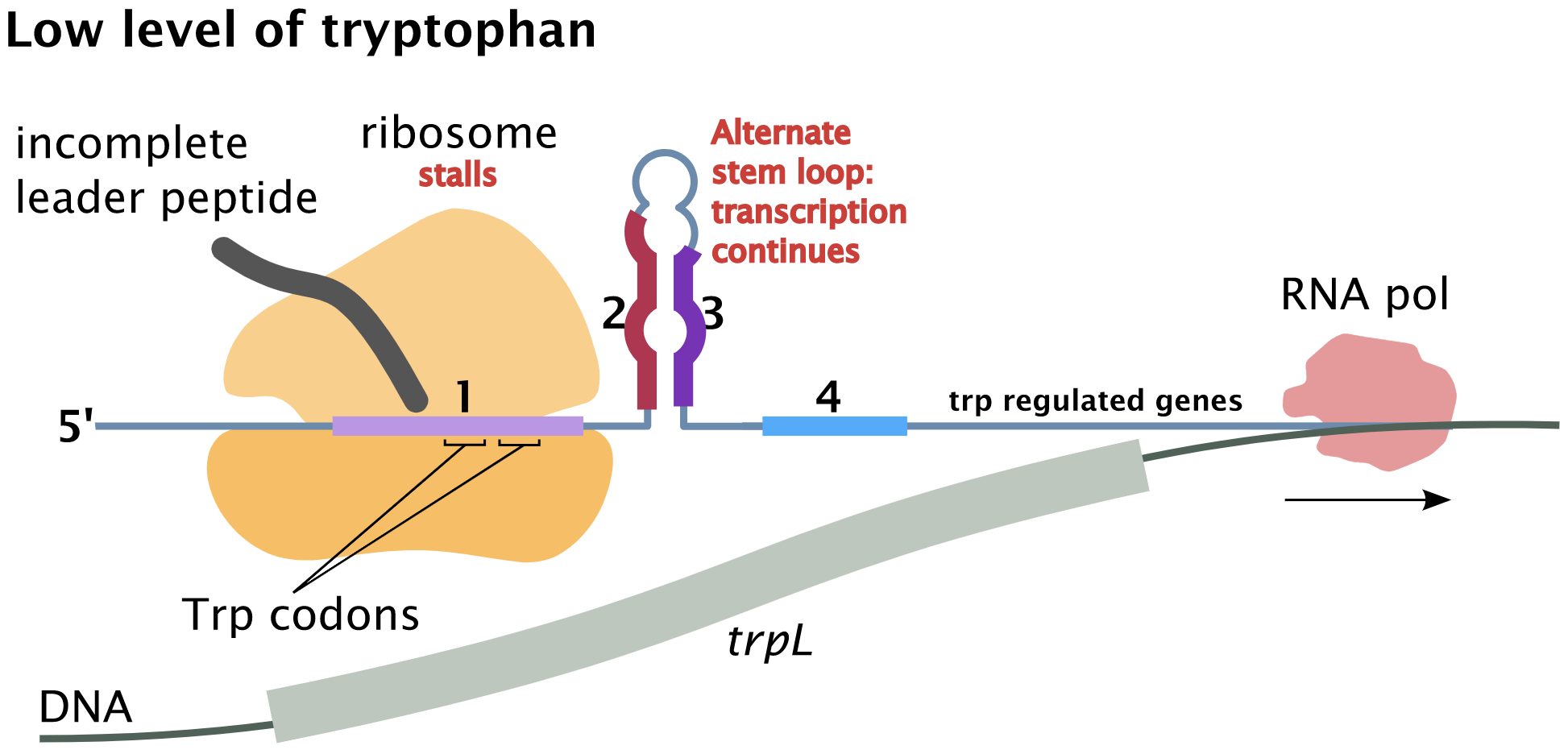}\includegraphics[width=.47\textwidth]{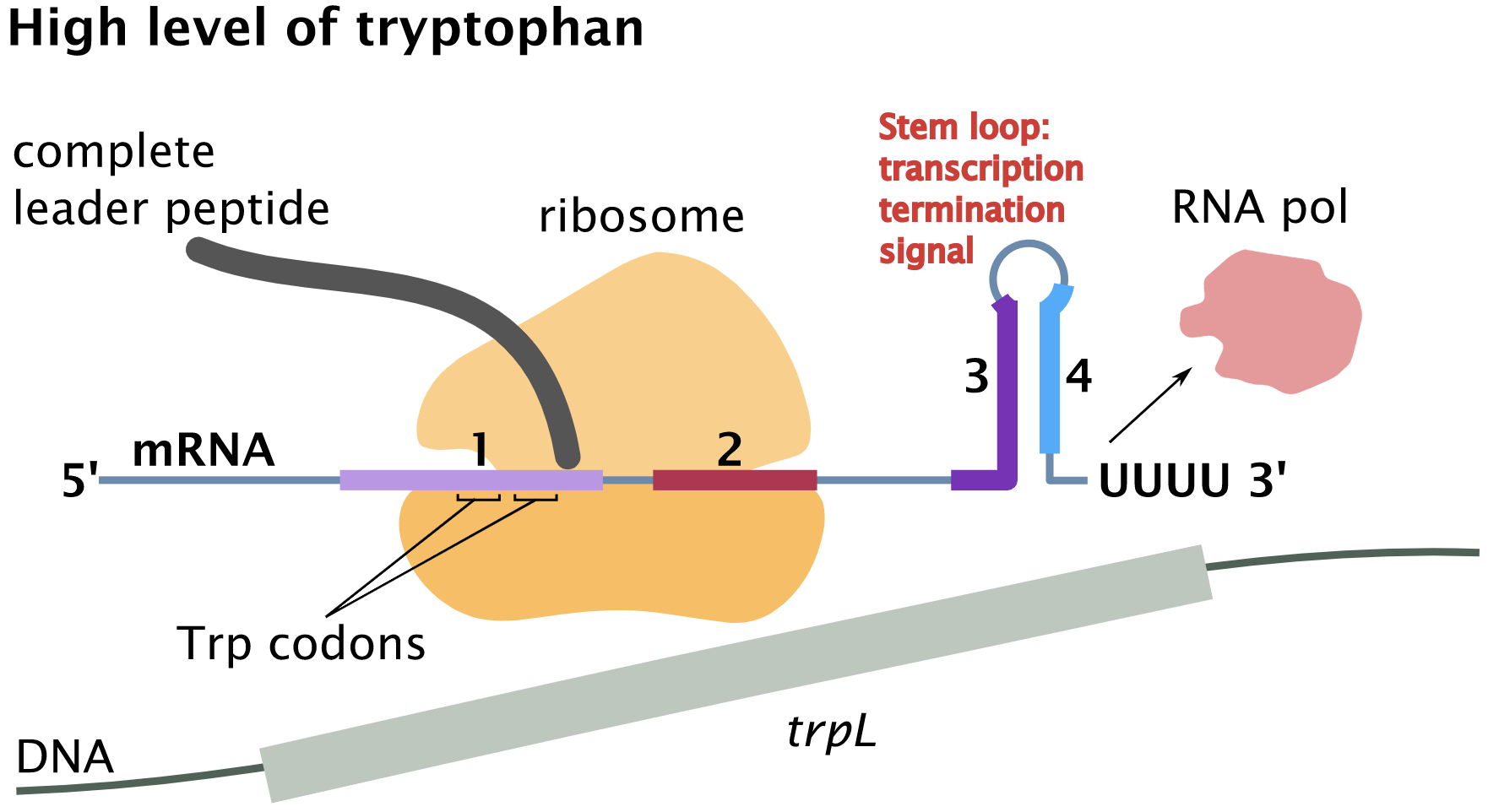}
\caption{Mechanism of attenuation. Translation of mRNA by ribosomes occurs while it is still being transcribed. Without charged tRNA, the ribosome stalls, the 2--3 hairpin will form (left), and a full-length mRNA sequence will be transcribed. Charged tRNA allows the ribosome to complete the leader peptide and fosters formation of the 3--4 hairpin loop (right), causing mRNA to be attenuated and resulting in only a short leader sequence.  Figure from Wikipedia, CC BY-SA 3.0.}\label{fig:trp-loop}
\end{figure}

The second type of regulation employed by the \emph{trp} operon is called \emph{attenuation}, and it can reduce the expression level by an additional factor of 10. Attenuation works because tryptophan binds to tRNA, and the presence or absence of these tryptophan-charged tRNA molecules can change how the mRNA is looped. Specifically, immediately upstream of the structural genes, in the \emph{leader} region, the mRNA contains four subsequences, labeled 1, 2, 3, and 4, each which is complementary to the next. As such, an adjacent pair of subsequences can bind to each other, which results in the mRNA forming a hairpin loop. Under low levels of tryptophan and therefore low levels of charged tryptophan tRNA, ribosomes pause at one of the two \emph{trp} codons in sequence 1, leading to the formation of the ``anti-termination'' hairpin loop formation between sequences 2 and 3, shown on the left in Figure~\ref{fig:trp-loop}. This allows the RNA polymerase to move forward and continue transcription. In contrast, when tryptophan is available to charge tRNA, the charged tRNA binds and the ribosome moves forward. The  ribosome blocks the formation of the 2-3 hairpin and increases the likelihood of the 3-4 ``termination'' hairpin forming, which is shown on the right in Figure~\ref{fig:trp-loop}. This causes the RNA polymerase to disassociate from the DNA strand, and transcription to terminate early, before reaching the structural genes. As a result, only a short mRNA ``leader'' sequence is transcribed, and none of the operon's proteins will be translated. The location of the leader region in the \emph{trp} operon, where attenuation is controlled, is shown in Figure~\ref{fig:trp-operon}.  

In \emph{E. coli}, mRNA is translated while it is still being transcribed. While the repressor protein is bound to the operon, it is impossible for transcription to occur, and if mRNA is attenuated, the translation of the five proteins cannot occur. However, neither of these repression methods occur with probability 1, regardless of how much tryptophan is in the cell. For example, under high levels of tryptophan, there is a high probability that any given TrpR molecule will be bound to tryptophan, and available to bind to the operator region to block transcription. However, there will be times when such bonds disassociate, and the repressor protein will fall off of the DNA strand, allowing transcription to begin, until another activated TrpR molecule reattaches. In other words, it is best to think about the \emph{probability} that a repressor protein is bound to the operon. No matter how much tryptophan is available, this probability is strictly less than $1$. Additionally, it will also be greater than 0, because even under conditions of extreme tryptophan starvation, there can always be occasional stray molecules present, as it is one of the fundamental amino acids.

Similarly, we should think of attenuation as a process that also happens with a probability strictly between $0$ and $1$. The assumption that there can always be trace amounts of tryptophan in the cell means that there can always be the occasional charged tRNA molecule as well. In other words, it is still possible for a single strand of mRNA to become attenuated under only trace levels of charged tRNA. On the other hand, even in an abundance of charged tRNA, there is still a nonzero probability that the 2--3 hairpin loop is occasionally formed, allowing non-attenuated mRNA to be transcribed. In this sense, the \emph{trp} operon can be thought of as a ``leaky faucet,'' which can never be fully turned off. As the theoretical limit of tryptophan concentration goes to infinity, the expression levels of the operon genes approach the \emph{basal level}, which we will denote by $b>0$. 

Since the repressor protein can decrease gene expression by a ballpark factor of $70$, and attenuation by another factor of $10$, we will assume that expression level of the \emph{trp} operon under extreme tryptophan starvation is in the neighborhood of $700b$, although some sources claim that under certain conditions, it can be over $1000b$ \cite{yanofsky1991physiological}. Our Boolean model will incorporate three expression levels: $700b$ will be considered ``high,'' and the basal level of $b$, which occurs under an abundance of tryptophan, is ``low.'' As tryptophan concentration decreases, the organism tries to preserve its ability to synthesize it, so any available tryptophan is more likely to be bound to tRNA than to the repressor protein. As a result, attenuation persists longer than the effects of the repressor protein \cite{yanofsky1984repression}. In our model, we will refer to this region as ``medium'', and the expression levels will be in the rough neighborhood of $10b$. 

The \emph{trp} operon uses a third method  of repression called \emph{feedback inhibition}, which occurs through tryptophan binding to anthranilate synthase, the initial protein in the tryptophan synthesis pathway. Anthranilate synthase is composed of two TrpE proteins and two TrpD proteins. Feedback inhibition of anthranilate synthase can reduce the production of tryptophan by a factor of $2$. Though this is an important function, it is primarily more of fine-tuning of self-regulation, and in our coarse-grained Boolean model, it will never change the concentration levels enough to jump between the low, medium, and high levels. For example, if feedback inhibition reduces the anthranilate synthase enzyme activity from $700b$ to $350b$, then this is still considered at a high level. Similarly, reducing from $10b$ to $5b$ just fine-tunes the concentration within the medium range. Because of this, our Boolean model will not incorporate feedback inhibition.

\subsection{Tryptophan metabolism and the \emph{tna} operon}

Whereas the \emph{trp} operon is used to synthesize tryptophan, the tryptophanase (\emph{tna}) operon is used to transport and metabolize it. The \emph{tna} operon is controlled by three regulatory mechanisms. The first is \emph{catabolite repression}, a system that prevents transcription from occurring if glucose, the cell's preferred carbon source, is available. The cAMP-CAP protein complex is needed to initiate transcription, much like how a key is needed to start an engine. The presence of glucose reduces cAMP levels, and hence depresses concentration levels of this transcription factor. Catabolite repression also occurs in the \emph{lac} and \emph{ara} operons, since glucose is preferred to those sugars as well. In all, the cAMP-CAP protein complex regulates several hundred genes in \emph{E. coli} \cite{zheng2004identification}. 

If tryptophan is available and glucose is not, then the cAMP-CAP protein complex binds to the promoter region allowing transcription to begin. The \emph{tna} operon includes two structural genes, \emph{tnaA} and \emph{tnaB}. The \emph{tnaA} gene encodes for TnaA, the tryptophanase enzyme used to metabolize tryptophan, and TnaB is a permease (transporter protein). This leads us to the second regulatory method, which involves the termination of transcription. There are two general ways that this is done in prokaryotes. The most common is called \emph{intrinsic termination}, which is carried out by mRNA forming a hairpin loop when a particular stop sequence is transcribed, causing the mRNA and the RNA polymerase to detach from the DNA and from each other. This is the mechanism of the attenuation described above. However, in the \emph{tna} operon, if tryptophan is not available, then transcription needs to be terminated immediately. This is carried out via a method called \emph{Rho dependent termination}, because it is meditated by the Rho ($\rho$) termination factor. Normally, this protein binds to mRNA when a cytosine-rich region is reached after the structural genes, and moves up the strand toward the RNA polymerase, removing it from the DNA strand. However, in the \emph{tna} operon, it attaches to the \emph{tnaC} leader sequence of the mRNA, which is downstream of the promoter region but upstream of the \emph{tnaA} and \emph{tnaB} genes. When tryptophan is present, the Rho protein is prevented from binding at this site. This occurs because tryptophan prevents the ribosome producing the TnaC peptide from releasing from the mRNA, thus blocking the Rho binding site and allowing the RNA polymerase to continue transcription of the \emph{tnaA} and \emph{tnaB} genes  \cite{gong2001mechanism}. This ensures that the operon remains on as long as there is tryptophan needing to be metabolized.

The Mtr and AroP permeases, which are not contolled by the \emph{tna} operon, are able to transport tryptophan into the cell, though not at the rate at which it can be transported if the operon is expressing the \emph{tnaB} gene. As such, we can assume that if tryptophan is available externally, some of it will make its way into the cell. This, in turn, decreases the probability that the Rho protein is able to bind to the mRNA. Which in turn, increases translation of the TnaA enzyme and TnaB transporter protein, which allows more tryptophan into the cell to be metabolized. This positive feedback loop increases the concentration of the gene products, until they approach a steady-state concentration. At this point, the operon is on. Eventually, when the external tryptophan is depleted, fewer gene products are translated, enzyme concentration levels decrease, and the probability that the Rho protein terminates transcription early increases. This turns the operon off. 

Additional regulatory mechanisms (one transcriptional, and one post-translational) were discovered and elucidated in 2014 and 2015 \cite{li2014camp-independent, li2015new}. Adding glucose to the cell in a way that retained the cAMP-CAP complex was shown to reduce expression levels of the \emph{tnaA} gene by approximately four-fold. Furthermore, the addition of glucose also reduces the activity of the tryptophanase enzyme already present in the cell. In summary, the  regulatory mechanisms cause the transcription of \emph{tna} mRNA to occur precisely when there is tryptophan available but no glucose. Assuming glucose is absent, the operon will be off if tryptophan levels are low, and on if tryptophan levels are high. Between these two extremes, the expression level is highly non-linear and sigmoidal as a function of the concentration of tryptophan, and so medium expression levels are not observed. Moreover, it has been hypothesized that the \emph{tna} operon can exhibit \emph{bistability}, which means that for a middle region of tryptophan concentration, the operon could be off or on, depending on whether tryptophan levels are increasing or decreasing. Specifically, if cells were raised in a tryptophan starved environment, then as tryptophan levels increase from low to medium levels, the operon will remain off until a certain ``up-threshold'' concentration is reached. However, the ``down-threshold,'' at which the operon turns itself off, is lower. Therefore, when the concentration of tryptophan is between these two levels, some cells will be found expressing the \emph{tna} operon's genes, and other will not. Like our Boolean model of the \emph{trp} operon, our model of the \emph{tna} operon will also consider three levels of tryptophan.

\subsection{Prior models}

After the \emph{lac} operon, the \emph{trp} operon is probably the most well-studied basic systems of gene regulation. Not long after it was discovered, Goodwin modeled it with a very simple set of ordinary differential equations (ODEs) \cite{goodwin1965oscillatory}. In 1982, Bliss et al. published a more complicated ODE model incorporating repression and feedback \cite{bliss1982role}, but this did not consider attenuation, which had only been discovered the previous year. However, their model did incorporate time-delays due to transcription and translation. Such a system of ODEs are called \emph{delay differential equations} (DDEs). The \emph{trp} operon has also been modeled in \cite{simao2005qualitative} with Petri nets, a graphical agent-based framework. In 2001, Santill\'{a}n and Mackey modeled the \emph{trp} operon with a system of DDEs \cite{santillan2001dynamic} that incorporated attenuation and more up-to-date and accurate knowledge about certain cellular processes. A review of this model and their DDE model of the \emph{lac} operon can be found in \cite{mackey2004modeling}.

In 2019, Santill\'{a}n  and his collaborators proposed the following ODE model for the \emph{tna} operon \cite{orozco2019bistable}:
\begin{align*}
    A'&=k_AP_G(G_e)P_W(W)-(\gamma_A+\mu)A \\
    B'&=k_BP_G(G_e)P_W(W)-(\gamma_B+\mu)B \\
    W'&=(\alpha+\beta B)W_e-(\delta+\epsilon A P_A(G_e,W_e)+\mu)W.
\end{align*}
Here, $A(t)$, $B(t)$, and $W(t)$ represent the concentrations of  tryptophanase (TnaA), the TnaB permease, and intracellular tryptophan, respectively. The concentrations of extracellular tryptophan and glucose are assumed to be parameters (constants), and are denoted $W_e$ and $G_e$, respectively. The rate constants $k_A$ and $k_B$ are from the assumption of mass-action kinetics. The terms $\gamma_AA$ and $\gamma_BB$ model protein degradation, and the terms $-\mu A$ and $-\mu B$ model dilution due to cellular growth. The function $P_G(G_e)$ is a sigmoidal function that accounts for catabolite repression. 

Since the \emph{tna} operon is not as well studied as many others, precise values of the parameters and rate constants are not known. Instead, the authors of \cite{orozco2019bistable} estimated the parameters of their model by fitting it to their experimental data where they exhibited bistability. After doing this, they varied the value of external tryptophan, $W_e$, the operon's inducer, and numerically computed the fixed points. They showed that their model has a unique fixed point if $W_e$ is low or high. If $G_e$ is not too high, then for certain mid-range values of $W_e$, their model has two stable fixed points separated by an unstable fixed point, which gives even more evidence that the \emph{tna} operon can exhibit bistability under the right conditions.

\section{A Boolean model of the \emph{trp} operon}\label{sec:trp}

\subsection{Model proposal and justification}

In this section, we will propose a Boolean model for the \emph{trp} operon. It will incorporate two regulatory processes that the \emph{trp} operon utilizes: repression and attenuation. As discussed above, enzyme feedback inhibition is more of a fine tuning mechanism, as it only depresses the levels by a factor of $2$. This is not significant enough to cause a jump from high to medium levels, or from medium to low, so it will not be incorporated in the model. 

Our model consists of eight Boolean variables and two parameters (constants). The variables represent quantities that can change rapidly. A value of $0$ can be thought of as ``high'' and $1$ as ``low'', and this could refer to the \emph{concentration} of a particular gene product, or  \emph{probability} of whether a protein is bound to tryptophan, or the probability of a given \emph{trp} mRNA strand being attenuated. In contrast, the parameters represent quantities that change over a much larger timescale.  Solutions to ODEs that model enzyme kinetics are typically highly non-linear, often well-modeled by ``S-shaped'' sigmoidal or Hill functions. Boolean step functions can be thought of as idealized limits of these functions, much like how the popular Dirac delta function is an idealized limit of a tall spike from a unit impulse force. For a survey of this, see \cite[Chapter 2]{robeva2013mathematical}.

The two parameters in our \emph{trp} model describe intracellular (pre-existing) tryptophan. For both the \emph{trp} and \emph{tna} operons, we want to be able to speak of three concentration levels of tryptophan: low, medium, and high. We can do this by introducing an additional variable and parameter. We will denote our parameters with lowercase Greek letters, and they will represent the following:
\begin{itemize}
    \item $\omega_i=$ intracellular tryptophan (high)
    \item $\omega_{im}=$ intracellular tryptophan (at least medium levels).
\end{itemize}
These parameters allow us to describe three levels of tryptophan in the cell: $(\omega_i,\omega_{im})=(1,1)$ means high levels, $(\omega_i,\omega_{im})=(0,1)$ means medium levels, and $(\omega_i,\omega_{im})=(0,0)$ describes low levels. The fourth possibility, of $(\omega_i,\omega_{im})=(1,0)$, is meaningless and can be ignored, much like how one disregards the region of the phase space of an ODE model where concentration or population is negative.   

We will use capital Roman letters to denote our eight variables, in a manner that suggests what they represent. As we did with the parameters for tryptophan, we will introduce an extra variable for tryptophan so we can capture three distinct levels: $(W,W_m)=(1,1)$ denotes high concentration of tryptophan, $(W,W_m)=(0,1)$ represents medium concentration, and $(W,W_m)=(0,0)$ means low concentration. The fourth possibility, $(W,W_m)=(1,0)$, can be ignored. We will often use variables of the form $x_i$ for convenience, so we can use notation like $(x_1,\dots,x_8)$. Both conventions are listed below.
\begin{itemize}
    \item $x_1=E=$ anthranilate synthase enzyme (high levels)
    \item $x_2=E_m=$ anthranilate synthase enzyme (at least medium levels)
    \item $x_3=L=$ the 3-4 hairpin loop is formed, causing \emph{trp} mRNA to be attenuated
    \item $x_4=M=$ \emph{trp} mRNA (either attenuated or not)
    \item $x_5=R=$ the TrpR repressor protein is activated
    \item $x_6=T=$ tRNA is charged
    \item $x_7=W=$ synthesized tryptophan (high levels)
    \item $x_8=W_m=$ synthesized tryptophan (at least medium levels).
\end{itemize}
The Boolean variables $L$, $M$, $R$, and $T$ should be thought of as a particular \emph{probability} of being high or low. For example, as the concentration of charged \emph{trp} tRNA increases, so does the probability that a given \emph{trp} mRNA strand will become attenuated, and thus unable to translate the operon's gene products. Saying that ``charged tRNA levels are low'' is equivalent to saying that the probability of a given tRNA molecule being charged is low. In this case, the probability of the 3-4 loop forming (and mRNA being attenuated) is low. In contrast, we will say that ``charged tRNA levels are high'' if the probability of a given tRNA molecule being charged is high, and this causes the probability of the 3-4 loop forming, and hence a given mRNA strand being attenuated, to be close to 1. 

Probabilities are preferred because even under high levels of tryptophan, an activated repressor protein could occasionally fall off of the operator region, leading to the initiation of transcription, and the translation of the proteins that the \emph{trp} mRNA codes for. In other words, the probability of the repressor protein being bound to the operon at any point in time is always strictly less than 1, and this leads to basal levels of mRNA and the operon's gene products. 

Unlike the parameters, which do not change over the course of the model, the states of our variables can change based on the states of the other variables and parameters. As such, we need to propose Boolean functions for them that represent how they interact with each other, and these are described below. The symbols $\And$ and $\Or$ denote logical AND and OR, respectively, and a bar over a variable denotes NOT. If $X$ is a variable, then we will denote its ``update function,'' that sends $X(t)$ to $X(t+1)$, as $f_X$. For example,
$f_X=Y\And\Not{Z}$ is short for $X(t+1)=Y(t)\And\Not{{Z(t)}}$, i.e., $X$ will be $1$ at the next time-step if and only if, at the current time-step, $Y=1$ and $Z=0$. The Boolean functions of our \emph{trp} model are listed below.

\begin{itemize}
    \item The anthranilate synthase enzyme complex will form from the translated proteins TrpE and TrpD and be maintained at high levels if \emph{trp} mRNA is present and not attenuated. Thus, the Boolean function is $f_E=M\And\Not{L}$.
    \item There will be at least medium levels of anthranilate synthase if \emph{trp} mRNA is available to be translated, or if there were high levels of it at the previous time step. Thus, the Boolean function is $f_{E_m}=M\Or E$. 
    \item If mRNA is transcribed and there are sufficiently high levels of charged tRNA, then there is a high probability of the 3-4 hairpin loop forming, causing \emph{trp} mRNA to be attenuated. The Boolean function is $f_L=M\And T$.
    \item There is a high probability of transcription of \emph{trp} mRNA if the repressor protein is inactivated. The Boolean function is $f_M=\Not{R}$.
    \item If there are high levels of tryptophan in the cell, then there is a high probability that a given molecule of the repressor protein will be bound to tryptophan, activating it so it is capable of blocking transcription. Under medium levels of tryptophan, attenuation is relieved later than repression, and so tryptophan is more likely to be found in charged tRNA than to be bound to the repressor protein \cite{yanofsky1984repression}. Note that the tryptophan is bound to the tRNA with covalent bonds while binding of tryptophan to the repressor is due to weaker, non-covalent bonds, so that the tryptophan will dissociate from the repressor as the intracellular concentration of tryptophan decreases. The Boolean function for the repressor protein is thus $f_R=W\Or\omega_i$.
    \item Charged tRNA will be available if there is some tryptophan present. The Boolean function is thus $f_T=\omega_{im}\Or W_m$.
    \item There will be high levels of tryptophan synthesized if \emph{trp} mRNA is being transcribed without any attenuation. The Boolean function is $f_W=M\And\Not{L}$. 
    \item There will be some synthesized tryptophan in the cell if \emph{trp} mRNA is present, or if there were high levels of synthesized tryptophan in the previous time step, because in just one timestep, it will not completely degrade or be utilized for other cellular processes. The Boolean function is thus $f_{W_m}=M\Or W$.
\end{itemize}

\begin{figure}[!ht]
  \tikzstyle{v} = [circle,fill=white,draw,inner sep=0pt, minimum size=5mm] 
  \tikzstyle{w} = [rectangle, fill=white,draw,inner sep=0pt, minimum size=5mm] 
  \tikzstyle{act} = [draw, -stealth]
  \begin{tikzpicture}[scale=1.7]
    \draw[fill=black!10, rounded corners] 
    (.2,-.4) rectangle (2.8,2.2);
    \node[v] (E) at (1.5,1.8) {\small $E$};
    \node[v] (L) at (1,1) {\small $L$};
    \node[v] (M) at (2,1) {\small $M$};
    \node[v] (T) at (.5,.2) {\small $T$};
    \node[v] (W) at (1.5,.2) {\small $W$};
    \node[v] (R) at (2.5,.2) {\small $R$};
    \draw[inhib] (R) to (M); 
    \draw[activ] (M) to (E);
    \draw[activ] (M) to (L);
    \draw[activ] (M) to (W);
    \draw[activ] (W) to (R);
    \draw[activ] (W) to (T);
    \draw[inhib] (L) to (E);
    \draw[inhib] (L) to (W);
    \draw[activ] (T) to (L);
    \begin{scope}[scale=.9,shift={(5,.85)}]
      \node[anchor=west] at (0,1.6) {$f_E=M\And \Not{L}$};
      \node[anchor=west] at (0,1.2) {$f_{E_m}=M\Or E$};
      \node[anchor=west] at (0,.8) {$f_L=M\And T$};
      \node[anchor=west] at (0,.4) {$f_M=\Not{R}$ };
	  \node[anchor=west] at (0,0) {$f_R=W\Or\omega_i$ };
      \node[anchor=west] at (0,-.4) {$f_T=W_m\Or\omega_{im}$ };
      \node[anchor=west] at (0,-.8) {$f_W=M\And\Not{L}$ };
      \node[anchor=west] at (0,-1.2) {$f_{W_m}=M\Or W$ };
      \draw[decorate,decoration={brace,amplitude=5pt}] 
      (0,-1.3) --  (0,1.7); 
    \end{scope}
  \end{tikzpicture}
  \caption{Our Boolean model of the \emph{trp} operon and its wiring diagram. The pairs of variables representing high and medium levels (e.g., $E$ and $E_m$) have been collapsed into one node each, as have the variables and parameters for tryptophan.}\label{fig:trp-wiring-diagram}
\end{figure}
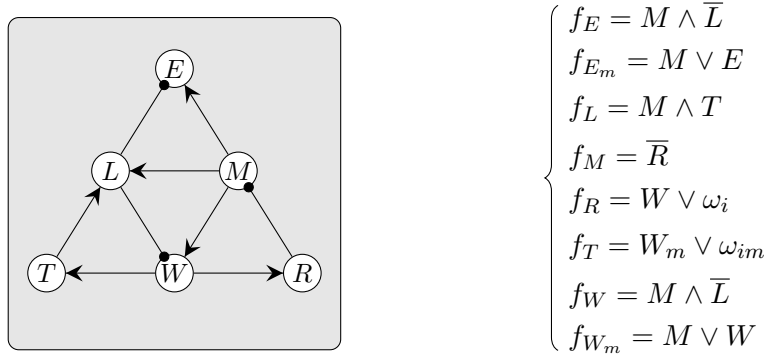

The relationship between the variables of the model can be encoded by a directed graph called the \emph{wiring diagram}, sometimes called the interaction or dependency graph, among others. The variables and parameters are represented as nodes, and an edge from $A$ to $B$ means that the state of $B$ is influenced by $A$. That is, the variable $A$ appears in the function $f_B$. If one wants to distinguish between a positive interaction, like an activation or transcription factor, and a negative interaction, like an inhibition or repression, then \emph{signed edges} can be used. There are numerous ways this can be done; we will use regular arrowheads for activations and circular arrowheads for inhibitions. A ``reduced'' wiring diagram of our model is shown in Figure~\ref{fig:trp-wiring-diagram}. Here, rather than having one node for each parameter and variable, we use one node for each molecular species, and collapse variables representing high and medium levels of the same molecule into a single node. Additionally, we represent intracellular tryptophan, whether it is a parameter or variable, with a single node. Notice that there are negative feedback loops (i.e., a feedback loop with an odd number of negative edges) consisting of $\{R,M,W\}$ as well as $\{T,L,W\}$, and a positive feedback loop consisting of $\{M,L,W,R\}$.

\subsection{Dynamics of the \emph{trp} Boolean model}\label{subsec:trp-dynamics}

As we have defined it, our Boolean model consists of eight functions based solely on the local interactions of the entities represented by the variables. Nowhere in the model have we specified any way to update the functions to generate dynamics. Every update scheme is going to be artificial, some more than others. However, certain salient features of a system are robust enough that they should be able to be captured even under manufactured dynamics. An example of this are the fixed points, which do not depend on whether the functions are updated synchronously or one of the many ways to update them asynchronously. We consider the dynamics of various update schemes to be part of the \emph{model validation} process, rather than the model itself. If observed dynamical features agree with what we expect to see biologicaly, then this provides strong evidence that our model has value. 

A standard way to update the functions of a Boolean model is synchronously, at regular discrete time steps. This generates a function
\[
f\colon\F_2^8\longto\F_2^8,\qquad
f\colon x\longmapsto (f_1(x),\dots,f_8(x)),
\]
where $\F_2=\{0,1\}$, and $x=(x_1,\dots,x_8)\in\F_2^8$ is a \emph{global state}, sometimes written as a binary string $x=x_1x_2\cdots x_8$ for brevity. The \emph{phase space} of our Boolean model has $2^8=256$ nodes, and every state will eventually end up in a periodic cycle or a fixed point. The set of states that feed into a periodic cycle is called a \emph{basin of attraction}. Naturally, such a synchronous update is undoubtedly artificial, and very little, if anything, can be deduced from the individual transitions. However, one of the most important features of a model is its long-term behavior, such as its fixed points, and these should be accurately represented.

We will now turn our attention to analyzing the long-term behavior of our Boolean model of the \emph{trp} operon, to verify that it makes sense biologically. We need to do this for all three choices of parameters, $(\omega_i,\omega_{im})=(1,1)$, $(0,1)$, and $(0,0)$. We will refer to these as the \emph{parameter vector}. In general, a Boolean model might have multiple connected components, and the limit cycle or fixed point reached can depend on the initial state. We should not expect such sensitivity in our Boolean model. Instead, there should be a unique fixed point reached that depends only on the parameter vector: if tryptophan levels are high, then the operon should remain off. If tryptophan levels are low or moderate, then the operon should remain on. Since our model only has $8$ nodes, it is small enough to simulate completely, and we did this with the BoolNet package \cite{mussel2010boolnet} in the statistical programming language of R. This package is freely available, well-documented, and easy to use. For each choice of parameter vector, we write the logical functions to a text file. The first line of this file must be ``\verb+targets, factor+'', and then the remaining lines contain the functions. As an example, the function ``$f(E)=M\And\Not{L}$'' is encoded with the line \verb+E, M & !L+. If the file for $(\omega_i,\omega_{im})=(0,0)$ is named \verb+trp00.txt+, then running the following commands in R will return the attractors under a synchronous update, plot the phase space (or state space) graph, and find the attractors in the asynchronous automaton, which we will describe soon.
{\small
\begin{verbatim}
    install.packages("BoolNet")
    library(BoolNet)
    trpNetwork00 <- loadNetwork("trp00.txt")
    getAttractors(trpNetwork00)
    plotStateGraph(getAttractors(trpNetwork00))
    getAttractors(trpNetwork00,type="asynchronous",startStates=256)
\end{verbatim}
}

\begin{figure}[!ht]
\includegraphics[width=3in]{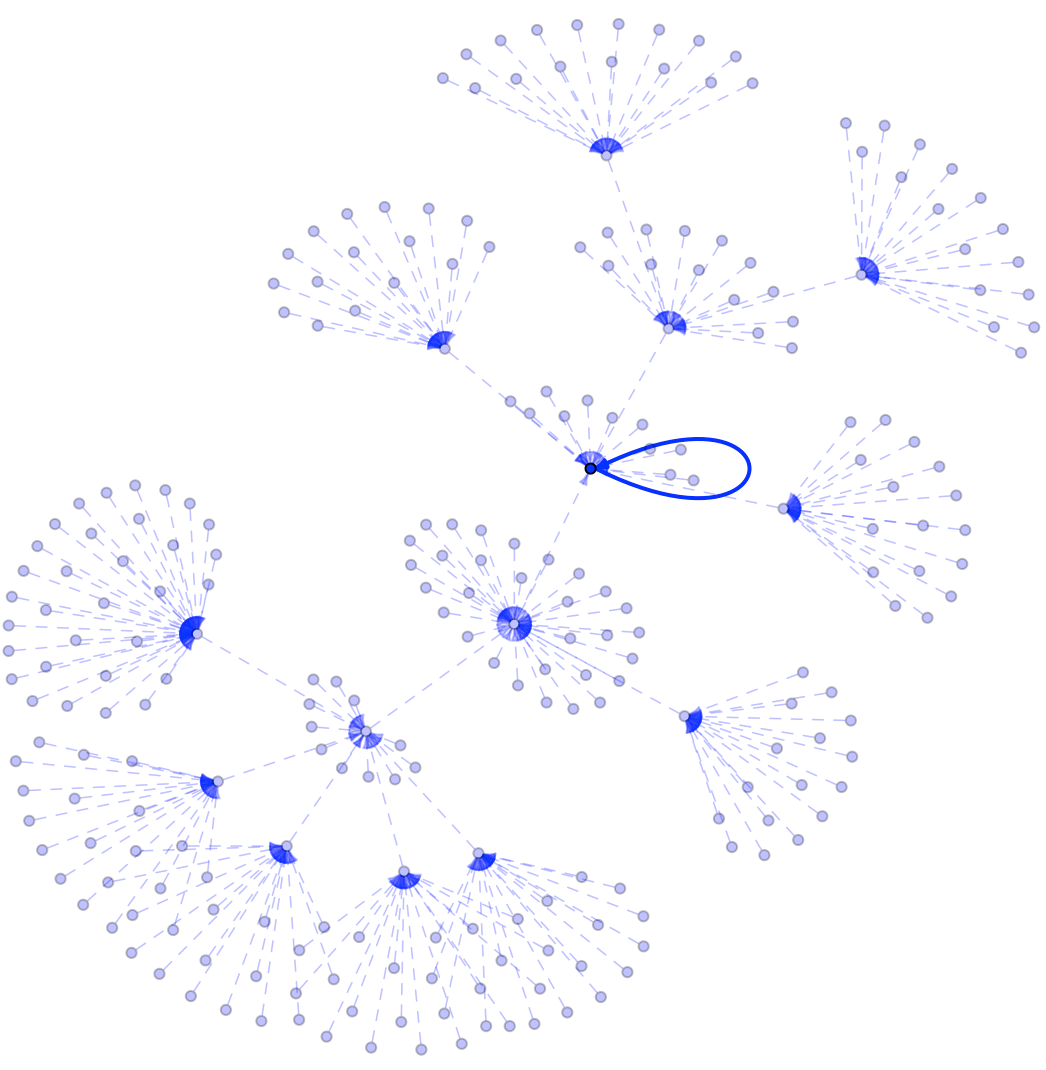}
\caption{Phase space graph of the \emph{trp} operon model with high tryptophan, $(\omega_i,\omega_{im})=(1,1)$. All of the initial states end up at the fixed point that we should expect biologically.}\label{fig:R_trp_high}
\end{figure}

Under high levels of tryptophan, i.e., $(\omega_i,\omega_{im})=(1,1)$, there is only one basin of attraction, and every state leads into the fixed point $x^*\!=x_1x_2\cdots x_8=00001100$. Let's briefly justify why this makes sense biologically.  If tryptophan levels are high, then it will bind to the repressor protein, activating it so it can bind to the operator region. As such, \emph{trp} mRNA will not be transcribed, so the proteins of the operon (TrpE, TrpD, etc.) will not be produced, and additional tryptophan will not be synthesized. In other words, $R=T=1$, and all other variables should be $0$. The phase space of the model is shown in Figure~\ref{fig:R_trp_high}.

The dynamics are a little more complicated for the other two parameter vectors, and initially, they appear potentially problematic. Let's first consider the case when there is some preexisting tryptophan present i.e., $(\omega_i,\omega_{im})=(0,1)$. At such medium levels, more of the available tryptophan will be bound to tRNA than to the repressor, as shown by Yanofsky, Kelley and Horn \cite{yanofsky1984repression}. Therefore, in this case, there will be an elevated probability of tRNA molecules being charged, but a lower probability of the repressor being bound to tryptophan. This will result in \emph{trp} mRNA being transcribed, but sometimes attenuated, and so anthranilate synthase will be synthesized, but not at the highest levels. Similarly, tryptophan will be synthesized, but also at medium levels. This would be described by the fixed point $x^*\!=01110101$, where $E=W=0$ means these molecules are not present at high levels, and $R=0$ because the repressor protein is not activated. 

This fixed point is indeed observed in the phase space, but it is only in a basin of attraction of size $8$. In other words, only $3.1\%$ of the global states lead into this fixed point. Another basin with $112$ states contains the $2$-cycle $01011101 \longleftrightarrow 11100111$, and the remaining $140$ states are in a basin that ends up in the $6$-cycle, defined by
\[
00010000 \longto 11010011 \longto 11011111 \longto 11101111 \longto 01001101 \longto 00000100. 
\]
These long cycles do not make sense biologically, and we will return to address them shorty, after discussing the case of low tryptophan, which is similar.  

\begin{figure}[!ht]
\includegraphics[width=3in]{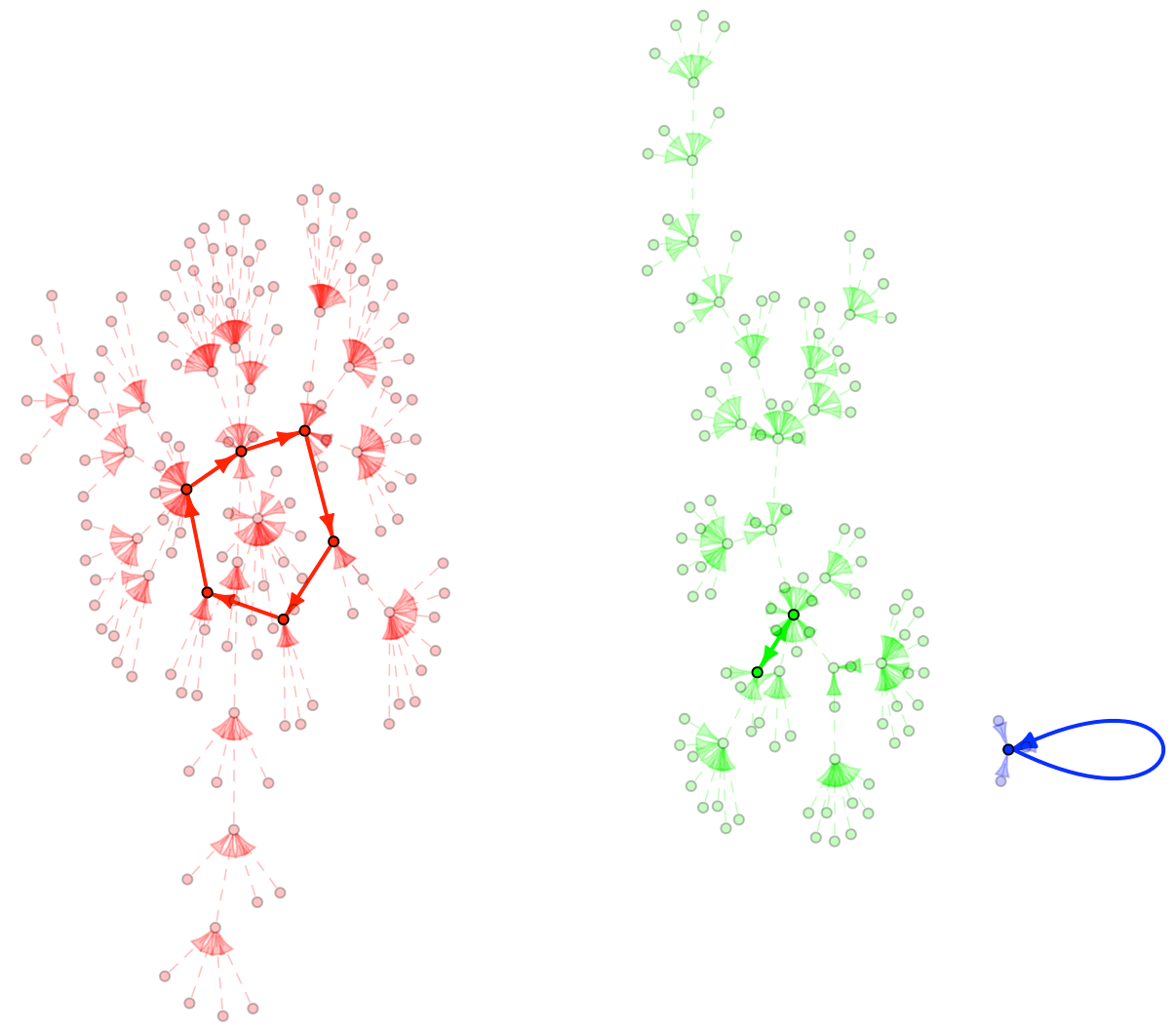}~\includegraphics[width=3in]{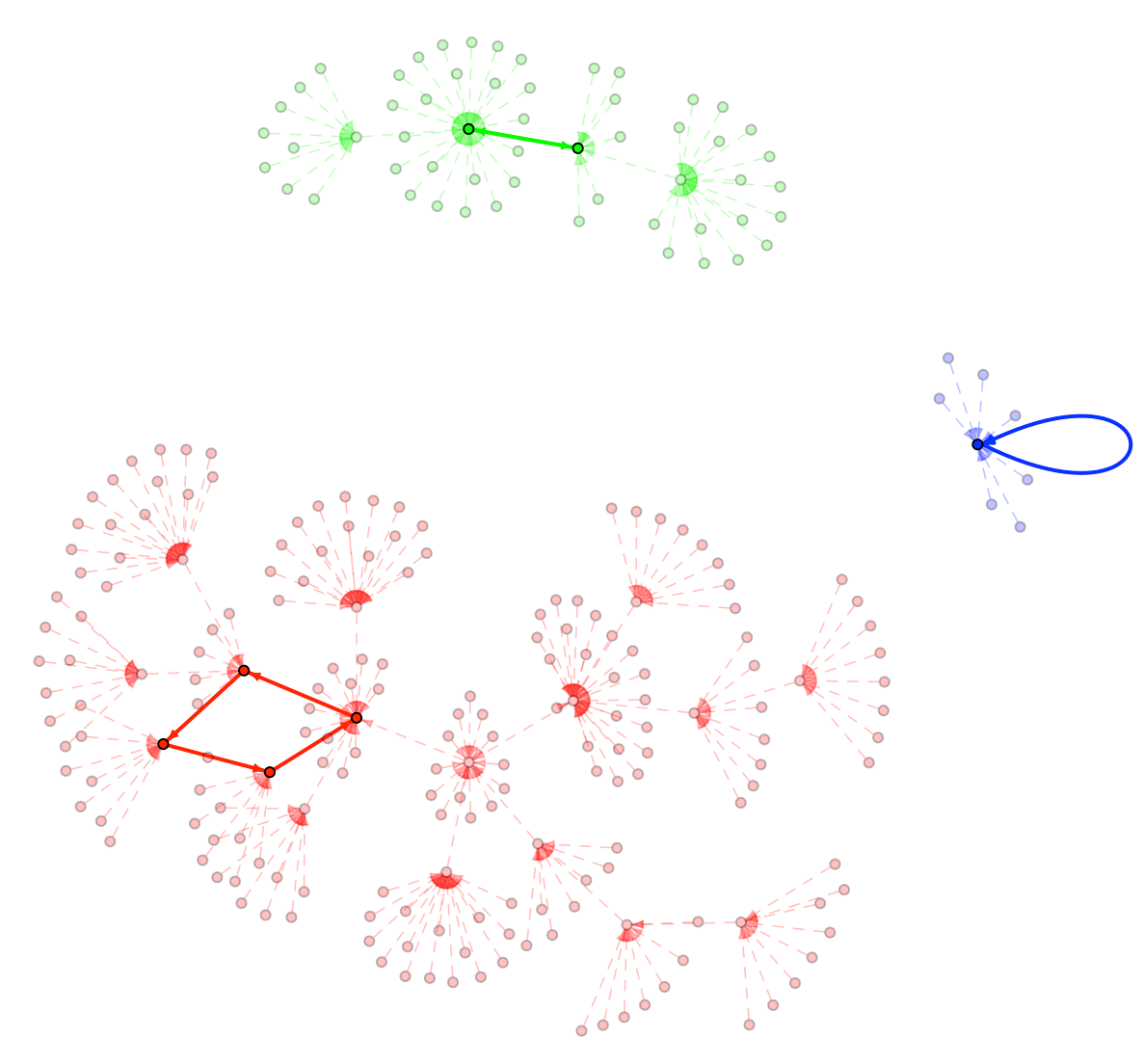}
\caption{The three components on the left make up the phase space of our \emph{trp} operon model with low tryptophan, $(\omega_i,\omega_{im})=(0,0)$. On the right is the phase space with medium tryptophan, $(\omega_i,\omega_{im})=(0,1)$. In both cases, the (blue) fixed point is what we should expect biologically.}\label{fig:R_trp_low-med}
\end{figure}

Under extreme tryptophan starvation, $(\omega_i,\omega_{im})=(0,0)$, the repressor protein remains unbound to tryptophan, and therefore, not able to block transcription. Transcription will begin, and tryptophan is not available to charge tRNA, and so the probability that \emph{trp} mRNA will be attenuated is very low. However, as the gene products are translated and tryptophan is synthesized, these molecules will charge tRNA, thereby increasing the likelihood of mRNA attenuation. As a result, the operon will remain on, but due to the nonzero probability of attenuation, both the anthranilate synthase enzyme and tryptophan will settle down to an equilibrium level below what they may have briefly appeared where there was no trypotophan in the environment. This also agrees with the same fixed point $x^*\!=01110101$ observed under medium levels of tryptophan. This global state is indeed a fixed point in the phase space, but this time, it is contained in a basin of attraction of size $4$. The same $2$-cycle $01011101 \longleftrightarrow 11100111$ exists, but in a basin of size $56$. The remaining $192$ states are in a basin with a $4$-cycle defined by
\[
00010100 \longto 11110111 \longto 01111101 \longto 01100101.
\]
The phase spaces of our Boolean model under these last two parameter vectors is shown in Figure~\ref{fig:R_trp_low-med}.

The good news thus far is that for all three parameter vectors $(\omega_i,\omega_{im})$ the phase space of our Boolean model updated synchronously has a unique fixed point, and the one that we expect to see. The concern is that there are two other limit cycles in large basins that cannot be explained biologically. Fortunately, in both cases, when then functions are updated asynchronously, these limit cycles disappear, and every initial state ends up at the correct fixed point. To elaborate on what we mean by this, every Boolean model $(f_1,\dots,f_n)$ has an \emph{asynchronous automaton} which describes all possible state transitions that result by updating one function at a time. Formally, this is a directed graph with vertex set $\F_2^n$, and $n$ directed edges from each $x\in\F_2^n$. Specifically, for for each $i=1\dots n$, there is a transition from $x=(x_1,\dots,x_n)$ to 
\[  
(x_1,\dots,x_i,f_i(x),x_{i+1},\dots,x_n)
\]  
that describes the result of updating the $i^{\rm th}$ function from $x$. Naturally, many of these edges are self-loops and can be ignored because they do not affect the long-term behavior. Consider a Boolean model where the functions are being updated randomly and asynchronously from a fixed global state $x\in\F_2^n$. This defines a walk on the asynchronous automaton, and with probability $1$, this will end up in a strongly connected component, called an \emph{attractor}, from which it can never leave. Attractors of size $1$ are just the fixed points, and it is easy to see that these are precisely the same fixed points that exist in the synchronous phase space. Unlike the longer limit cycles in the synchronous phase space, longer attractors need not just be simple cycles. There can be more attractors in the asychronous automaton than in the synchronous phase space, or fewer. In our case, under both parameter vectors $(\omega_i,\omega_{im})=(0,1)$ and $(0,0)$, the three connected components of the synchronous phase space merge into a single component with a unique strongly connected component -- its fixed point. This is easily verified using the BoolNet package in R, though unlike the synchronous phase space, there is not an easy way to visualize the asynchronous automaton. In other words, if we relax the unnatural condition that the Boolean functions are updated synchronously, then all global states will end up at the unique fixed point that we expect to see biologically. We say that the longer limit cycles are an \emph{artifact of synchrony}, rather than inherent to the model. In retrospect, it was not necessary to compute the synchronous phase space in this process, but it is nonetheless an interesting observation to see the difference in dynamics between a synchronous and asynchronous update, and we will return to this topic at the end of this paper. Interestingly, the Boolean model of the arabinose operon also exhibited longer limit cycles that disappeared under an asynchronous update \cite{jenkins2017bistability}, whereas the Boolean model of the lactose operon from \cite{veliz-cuba2011boolean} did not have any longer limit cycles under a synchronous update. In \cite{jenkins2017bistability}, this was partially attributed to the fact that the \emph{lac} operon utilizes a form of repression called \emph{inducer exclusion} by glucose, whereas the \emph{ara} operon does not. The summary of the fixed points of our Boolean model of the \emph{trp} operon is shown in Table~\ref{tbl:trp-fixed-points}. 

\begin{table}[!ht]
\begin{tabular}{|c|c|c|c|}
\hline Tryptophan & Parameter vector & Fixed point & Operon \\ 
concentration & $(\omega_i,\omega_{im})$ & 
$\;(E,E_m,L,M,R,T,W,W_m)$ & state \\ \hline 
high & $(1,1)$ & $(0\,,\;0\,,\,0\,,\,0\,,\,1\,,\,1\,,\,0\,,\,0)$ & OFF \\ \hline
medium & $(0,1)$ & $(0\,,\;1\,,\,1\,,\,1\,,\,0\,,\,1\,,\,0\,,\,1)$ & ON \\ \hline 
low & $(0,0)$ & $(0\,,\;1\,,\,1\,,\,1\,,\,0\,,\,1\,,\,0\,,\,1)$ & ON \\ \hline 
\end{tabular}
\caption{Parameter vectors of our \emph{trp} operon Boolean model, and the corresponding fixed points.} \label{tbl:trp-fixed-points}
\end{table}

\section{A Boolean model of the \emph{tna} operon}\label{sec:tna}

\subsection{Model proposal and justification}

In this section, we will propose a Boolean model of the \emph{tna} operon. This time, extracellular tryptophan will be a parameter, and intracellular tryptophan a variable. Our model, like the ODE model in \cite{orozco2019bistable}, will end up having two fixed points under medium levels of tryptophan and the absence of glucose, providing further evidence for the operon's hypothesized bistability. This also provides a third example, along with published Boolean models of the \emph{lac} \cite{veliz-cuba2011boolean} and \emph{ara} operons \cite{jenkins2017bistability}, of how a coarse-grained Boolean model can capture a fundamental biological phenomenon such as bistability.  

The variables in our model represent \emph{tna} mRNA, the Rho termination protein, the products TnaA (tryptophanase) and TnaB (tryptophan permease) of the operon's two structural genes, the cAMP--CAP protein complex that ensures that transcription only begins in the absense of glucose, and intracellular tryptophan. Since bistability is characterized by having two stable steady states under medium levels of the inducer, our model needs to be able to incorporate three concentration levels of tryptophan: low, medium, and high. We will do this just like we did for the \emph{trp} operon, by using two Boolean variables for tryptophan, $W$ and $W_m$, to represent high concentration and (at least) medium concentration, respectively. Our proposed Boolean model of the \emph{tna} operon consists of the following seven variables.

\begin{itemize}
    \item $x_1=A=$ TnaA enzyme (tryptophanse) that metabolizes tryptophan
    \item $x_2=B=$ TnaB permease, the primary tryptophan transporter
    \item $x_3=C=$ cAMP--CAP protein complex that initiates transcription
    \item $x_4=M=$ \emph{tna} mRNA
    \item $x_5=P=$ the \emph{tnaC} leader sequence is bound to the Rho protein
    \item $x_6=W=$ intracellular tryptophan (elevated levels)
    \item $x_7=W_m=$ intracellular tryptophan.
\end{itemize}

As we did with the \emph{trp} operon, we chose these variables because the concentrations of the molecules they represent can change rapidly. For most of these, a value of $0$ means low concentration and $1$ means high concentration, or at least, significantly above basal levels. The exception, other than $W$ and $W_m$, which we already discussed, is the Rho protein, which we will assume is always present. The value $P=1$ represents a very high (approximately $1$) probability of Rho molecules being bound to the \emph{tnaC} leader sequence, and therefore mRNA should almost always terminate early, upstream of the structural genes. In contrast, $P=0$ represents a much wider range of possibilities, in all of which the \emph{tnaC} leader sequence can be sometimes found unbound to Rho. As an analogy, imagine that the Rho protein is a water sealant for a room that contains papers and electronics. It is successful at keeping things dry if it is 100\% effective, or close to it, in which case we say that $P=1$. If it leaks, whether it lets in 15\% or 85\% of the water, then the contents will be destroyed, and we say $P=0$. Going back to the Rho protein, if the \emph{tnaC} leader sequence can be sometimes found unbound to Rho, whether with probability 0.15 or 0.85, then some mRNA will be transcribed, and concentrations of it and the gene products will be above basal levels. 

Unlike the quantities represented by the variables, the concentrations of extracellular tryptophan and glucose, both which are involved in the \emph{tna} operon, change on a much larger timescale. We will treat concentrations of these molecules as constants, and represent them as parameters in the model. Once again, we will use Greek letters to denote them, and they are the following.

\begin{itemize}
    \item $\gamma=$ glucose
    \item $\omega_e=$ extracellular tryptophan (high levels)
    \item $\omega_{em}=$ extracellular tryptophan (at least medium levels)
\end{itemize}

We will make the blanket assumption that at least one glucose permease is available. In other words, if glucose is available in the medium, then it will be transported into the cell. We will also assume that the permeases Mtr and AroP are available to transport tryptophan into the cell, but TnaB is needed to maximize tryptophan content because it is the primary tryptophan transporter \cite{gu2013knocking}. For each of the seven variables above, we need to propose a Boolean function that represents how its state is influenced by the other variables and parameters.  

\begin{itemize}
     \item Tryptophanase (TnaA) is coded for by one of the two structural genes of the \emph{tna} operon, so it will only be translated if \emph{tna} mRNA is transcribed. Additionally, glucose causes the expression level of the \emph{tnaA} gene to rapidly decrease, using both transcriptional and post-translational mechanisms. 
     Therefore, the Boolean function for this is $f_A=M\And\Not{\gamma}$.
    \item The TnaB permease is coded for by the second structural gene of the \emph{tna} operon. It will also be translated if \emph{tna} mRNA is transcribed, so its function is $f_B=M$.
    \item Glucose reduces the levels of cAMP, which in turn prevents the formation of the cAMP--CAP protein complex, which is needed for transcription to initiate. The Boolean function for the cAMP-CAP complex is thus $f_C=\Not{\gamma}$.
    \item For \emph{tna} mRNA to be expressed, the cAMP-CAP protein complex must be available, and the \emph{tnaC} leader sequence must be at times unbound to the Rho repressor protein. The Boolean function is thus $f_M=C\And\Not{P}$.
    \item The probability of finding Rho repressor protein molecules bound to the \emph{tnaC} leader sequence is close to $1$ if tryptophan is not present. Therefore, its Boolean function is  $f_P=\Not{W}\And\Not{W_m}$.
    \item Tryptophan will be present inside the cell at elevated levels if extracellular tryptophan is available, and the TnaB permease is available to transport it. The Boolean function is thus $f_W=\omega_e\And B$. 
    \item There are three ways that there can be at least some tryptophan available in the cell. First, there can be some tryptophan outside of the cell and availability of the TnaB permease. Alternatively, there can be high levels of tryptophan outside the cell. Even under basal levels of the TnaB protein, some of it will be transported into the cell by the Mtr and AroP permeases. Finally, if there are already elevated levels of tryptophan inside the cell, then we may assume that it will not completely degrade or be utilized by other cellular functions by the next time-step. The Boolean function for $W_m$ is thus $f_{W_m}=(\omega_{em}\And B)\Or \omega_e\Or W$. 
\end{itemize}

\begin{figure}[!ht]
  \tikzstyle{v} = [circle,fill=white,draw,inner sep=0pt, minimum size=5mm] 
  \tikzstyle{w} = [rectangle, fill=white,draw,inner sep=0pt, minimum size=5mm] 
  \tikzstyle{act} = [draw, -stealth]
  \begin{tikzpicture}[scale=1.7]
   \begin{scope}[shift={(0,1)}]
    \draw[fill=black!10, rounded corners] 
    (-1.3,-.3) rectangle (1,2.3);
    \node[v] (W) at (0,2) {\small $W$};
    \node[v] (A) at (-.75,1) {\small $A$};
    \node[v] (P) at (0,1) {\small $P$};
    \node[v] (B) at (.75,1) {\small $B$};
    \node[v] (C) at (-.75,0) {\small $C$};
    \node[v] (M) at (0,0) {\small $M$};
    \node[w] (We) at (-1.75,2) {\small $\omega_e$};
    \node[w] (Ge) at (-1.3,.5) {\small $\gamma$};
    \draw[activ] (We) to (W); 
    \draw[inhib] (W) to (P);
    \draw[inhib,dashed] (A) to (W);
    \draw[activ] (B) to (W);
    \draw[activ] (C) to (M);
    \draw[inhib] (P) to (M);
    \draw[activ] (M) to (A);
    \draw[activ] (M) to (B);
    \draw[inhib] (Ge) to (A);
    \draw[inhib] (Ge) to (C);
    \end{scope}
    \begin{scope}[scale=.9,shift={(3.05,2.2)}]
      \node[anchor=west] at (0,1.2) {$f_A=M\And\Not{\gamma}$};
      \node[anchor=west] at (0,.8) {$f_B=M$};
      \node[anchor=west] at (0,.4) {$f_C=\Not{\gamma}$};
      \node[anchor=west] at (0,0) {$f_M=C\And\Not{P}$};
	  \node[anchor=west] at (0,-.4) {$f_P=\Not{W}\And\Not{W_m}$};
      \node[anchor=west] at (0,-.8) {$f_W=\omega_e\And B$};
      \node[anchor=west] at (0,-1.2) {$f_{W_m}=(\omega_{em}\And B)\Or \omega_e\Or W$};
      \draw[decorate,decoration={brace,amplitude=5pt}] 
      (0,-1.3) --  (0,1.3); 
    \end{scope}
  \end{tikzpicture}
  \caption{Our Boolean model of the \emph{tna} operon and its wiring diagram. The pairs of variables representing high and medium levels (e.g., $E$ and $E_m$) have been collapsed into one node each, as have the variables and parameters for tryptophan.}\label{fig:tna-wiring-diagram}
\end{figure}
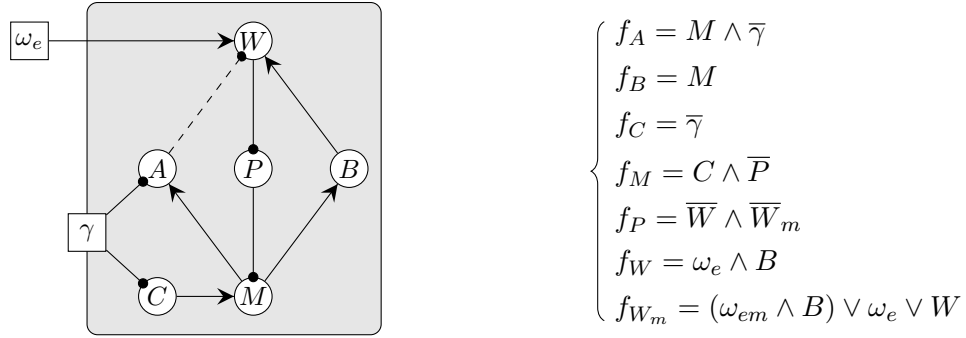

A wiring diagram of our model of the \emph{tna} operon is shown in Figure~\ref{fig:tna-wiring-diagram}. Once again, we collapse variables representing high and medium levels of the same quantity into a single node. The shaded region represents the cell. The parameters for extracellular tryptophan and glucose are boxed, as to distinguish them from variables, which are circled. The node for glucose is on the boundary of the cell, because of our blanket assumption that a glucose permease is available. In other words, if glucose is present, it will be found both in the medium, as well as inside of the cell. We included a dashed edge from $A$ to $W$ because even though TnaA metabolizes tryptophan, this effect is overshadowed by the presence of extracellular tryptophan entering the cell. Specifically, if tryptophan levels are high, and TnaB is available, then even if tryptophanse is around to metabolize it, more will enter the cell and replace what was consumed. Similarly, if there are only mid-range levels of tryptophan in the cell, then for it to persist, more extracellular tryptophan must enter the cell, regardless of how much is metabolized by TnaA. In other words, $A$ does not explicitly appear in the functions $f_W$ or $f_{W_m}$, but it does play a fine-tuning role in reducing tryptophan levels.

Notice how this wiring diagram has a positive feedback loop $\{M,B,W,P\}$, and a negative feedback loop $\{M,A,W,P\}$ if the dashed edge is included. A general rule of thumb in systems biology is that positive feedback loops are needed for multistability (multiple steady states), and negative feedback loops are needed for longer limit cycles~\cite{thomas1990biological}. Though this does not confirm that there is bistability, it is nonetheless something that we should expect. 

\subsection{Dynamics of the \emph{tna} model}

As we did with the \emph{trp} operon, we used the BoolNet package in R to simulate the dynamics of our \emph{tna} Boolean model under both a synchronous and asynchronous update. This time, there are three parameters, and the system needs to be analyzed separately for all possible combinations of the parameter vector $(\gamma,\omega_e,\omega_{em})\in\F_2^3$, except for the two where $\omega_e=1$ and $\omega_{em}=0$, because they are biologically meaningless. This leaves six cases, and we will begin with the three that describe the cell under the presence of glucose. In all of these cases, the operon should be off, so most variables will be zero. If extracellular tryptophan is high, i.e., $(\gamma,\omega_e,\omega_{em})=(1,1,1)$, then the Mtr and AroP permeases will bring some tryptophan into the cell, but the levels  will not be as elevated as if the primary transporter TnaB is available. It thus makes sense to expect to have $W_m=1$ in the long term. All other variables should be zero, leading to the fixed point $x^*\!=x_1x_2\cdots x_7=0000001$. For the other two parameter vectors with $\gamma=1$, not enough tryptophan will enter the cell to cause the Rho protein to detach from the \emph{tnaC} leader sequence. Therefore, all variables should be zero except for $P=1$, which describes the fixed point $x^*\!=0000100$.

\begin{figure}[!ht]
\includegraphics[width=1.9in]{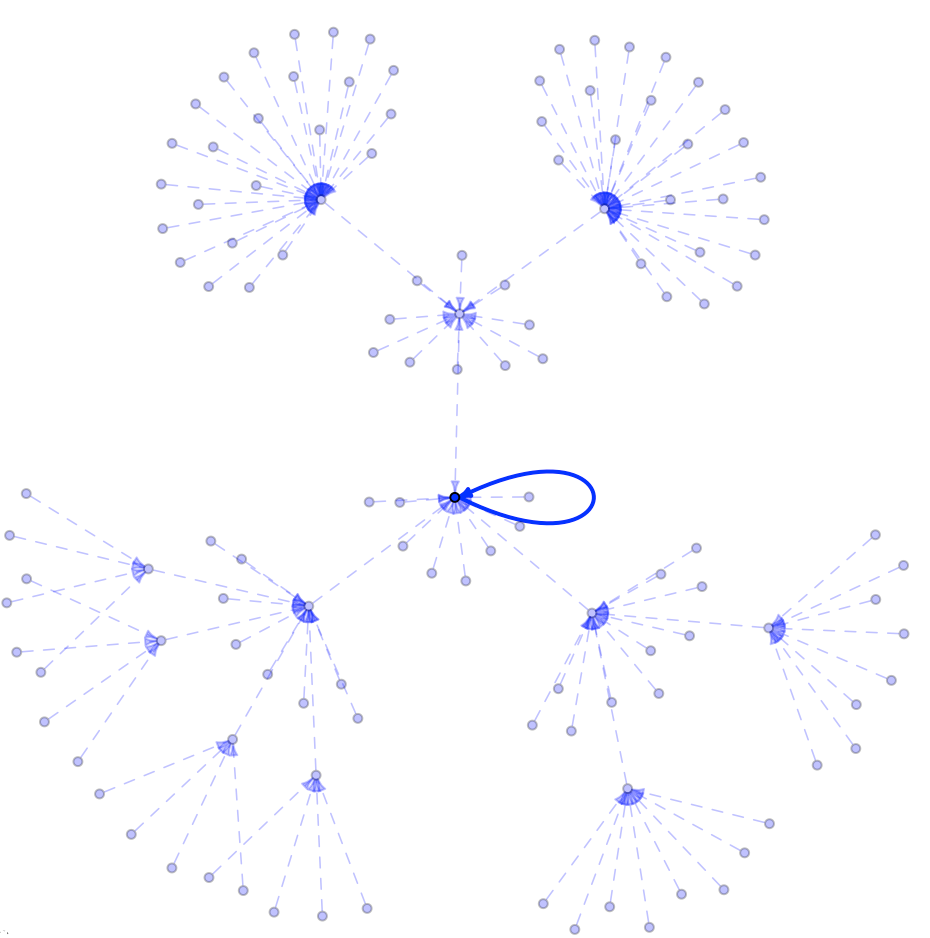}~\hspace{2mm}~\includegraphics[width=1.9in]{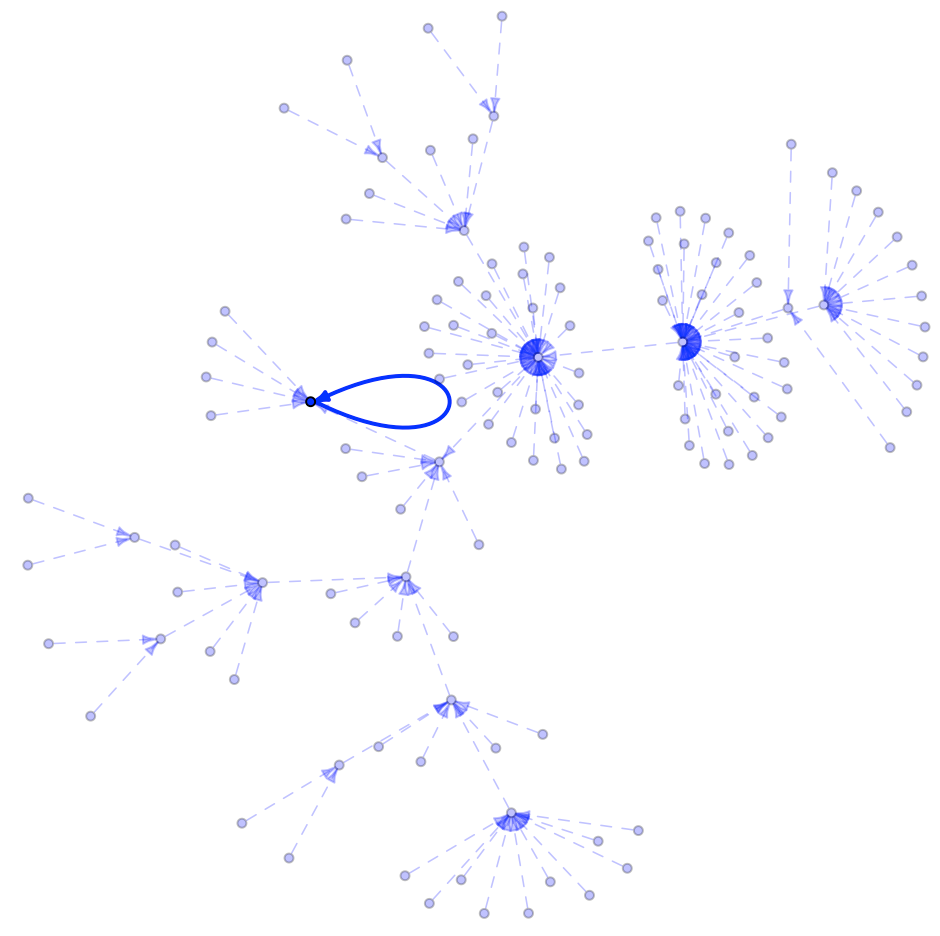}~\hspace{2mm}\includegraphics[width=2in]{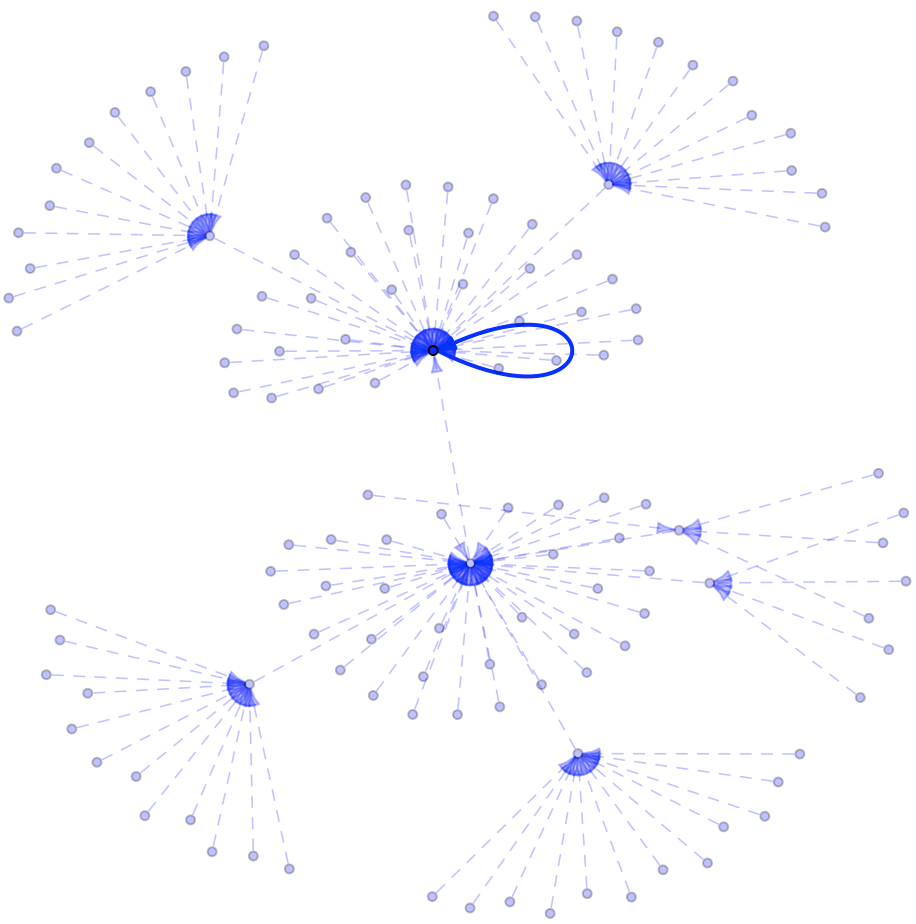}
\caption{Phase spaces of our \emph{tna} operon Boolean model with glucose: $\gamma=1$. From left to right are the cases where external tryptophan is low, medium and high. That is, $(\omega_e,\omega_{em})=(0,0)$, $(0,1)$, and $(1,1)$. In all three cases, the (blue) fixed point is what we should expect biologically.}\label{fig:R_tna_low-high_glu}
\end{figure}

In all three of the cases with $\gamma=1$, the dynamics that result from our Boolean functions updated synchronously have a unique basin of attraction containing precisely the fixed point that we expect to see, described above. The phase space graphs, as rendered by BoolNet, are shown in Figure~\ref{fig:R_tna_low-high_glu}.

Next, let's consider the cases when glucose, the cell's preferred carbon source, is not available. Starting with the parameter vector $(\gamma,\omega_e,\omega_{em})=(0,0,0)$, it makes sense that the corresponding fixed point describes the operon being off, since no tryptophan is available to be metabolized. Naturally, $C=1$ because the cAMP-CAP protein complex will be present in the absence of glucose, and $P=1$ because there is a high probability of the Rho protein being bound to the \emph{tnaC} leader sequence, causing termination of transcription. This is described by the fixed point $x^*\!=0010100$. 

When the parameter vector is $(\gamma,\omega_e,\omega_{em})=(0,1,1)$, there is no glucose and high levels of tryptophan, and so the operon should be on. Once again, the absence of glucose explains the presence of the cAMP-CAP protein complex $(C=1)$. The presence of the gene products is guaranteed by $A=B=1$, translated from \emph{tna} mRNA ($M=1$), which is transcribed because the  \emph{tnaC} leader sequence is most likely to be found unbound to the Rho protein ($P=0$). Naturally, there will be high levels of tryptophan, and so $W=W_m=1$. In other words, this is described by the fixed point $x^*\!=1111011$. 

\begin{figure}[!ht]
\includegraphics[width=2.75in]{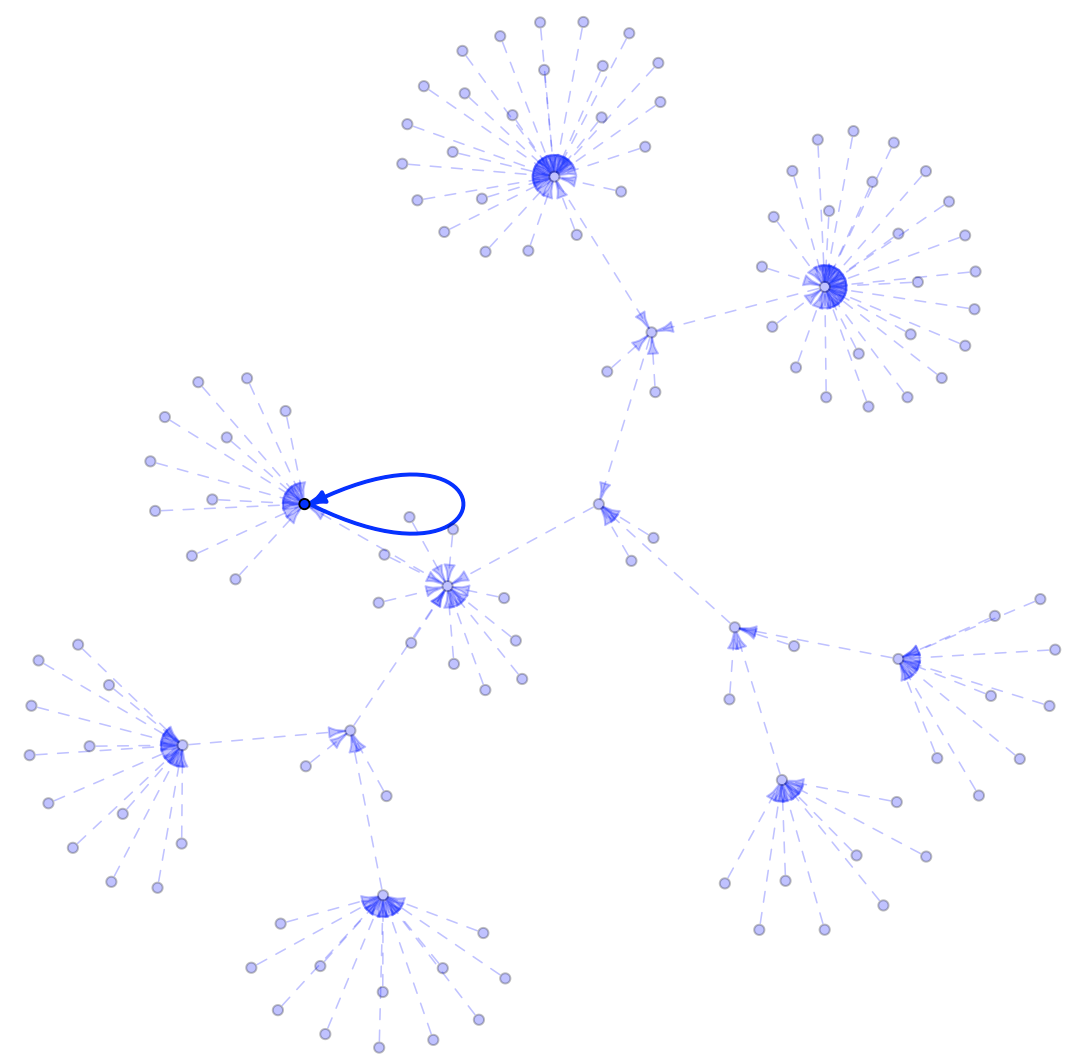}~\hspace{5mm}~\includegraphics[width=2.75in]{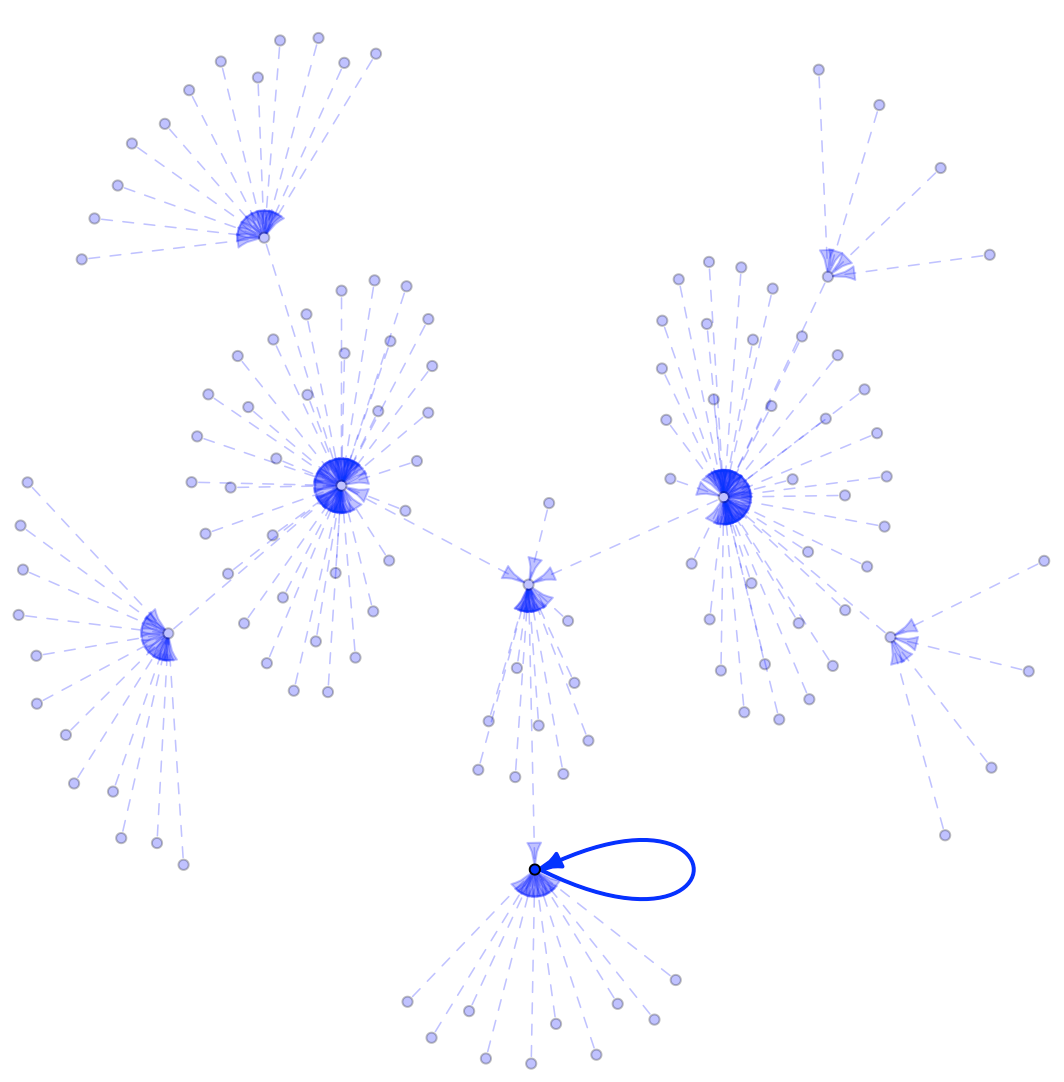}
\caption{Phase spaces of our \emph{tna} operon Boolean model with no glucose: $\gamma=0$. At left is the case of low external tryptophan, $(\omega_e,\omega_{em}=(0,0)$, and at right is high tryptophan, $(\omega_e,\omega_{em}=(0,0)$. In both cases, the (blue) fixed point is what we should expect biologically.}\label{fig:R_tna_low-high_noglu}
\end{figure}

In both of the aforementioned cases, our Boolean models updated synchronously defines a phase space with a unique basin of attraction leading into the fixed point that we expect. Figure~\ref{fig:R_tna_low-high_noglu} shows the dynamics as rendered by BoolNet. 

\begin{figure}[!ht]
\includegraphics[width=3.25in]{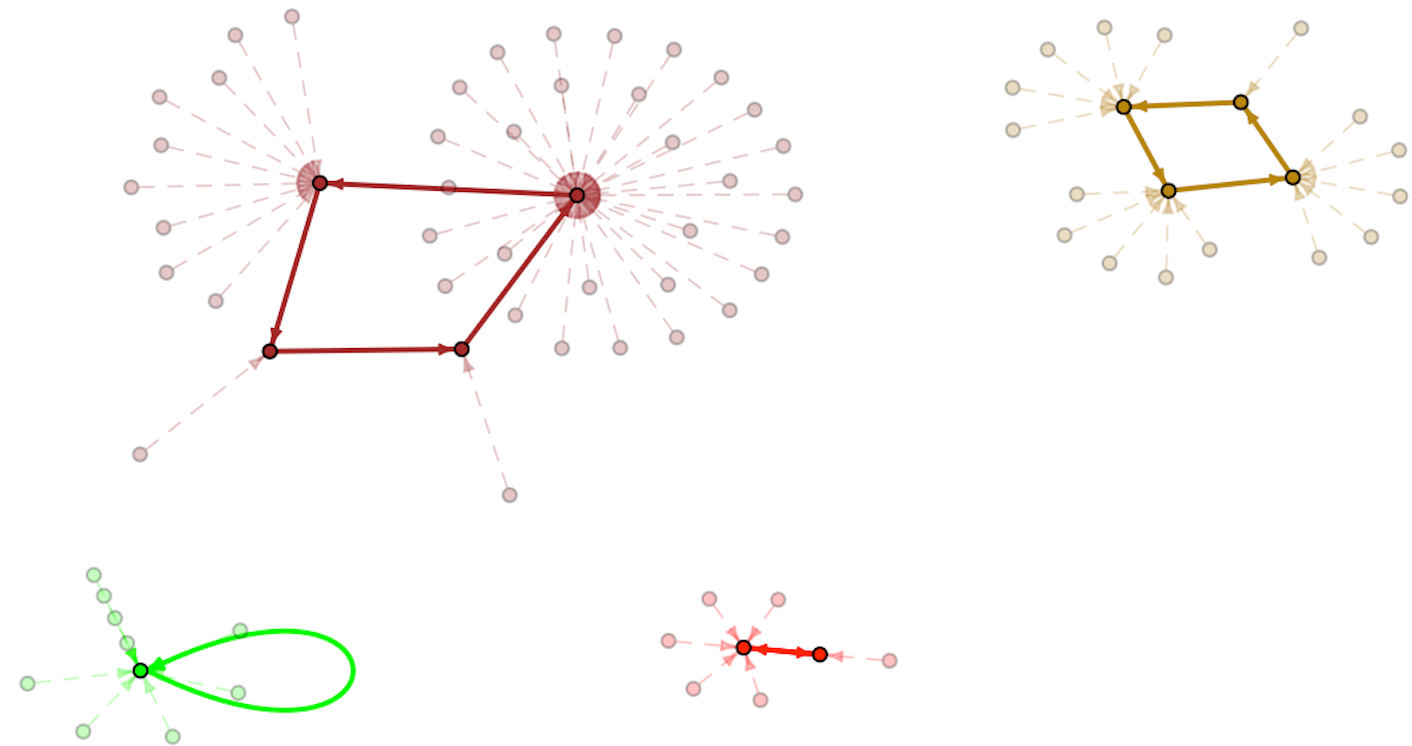}\hspace{-14mm}
\includegraphics[width=2.75in]{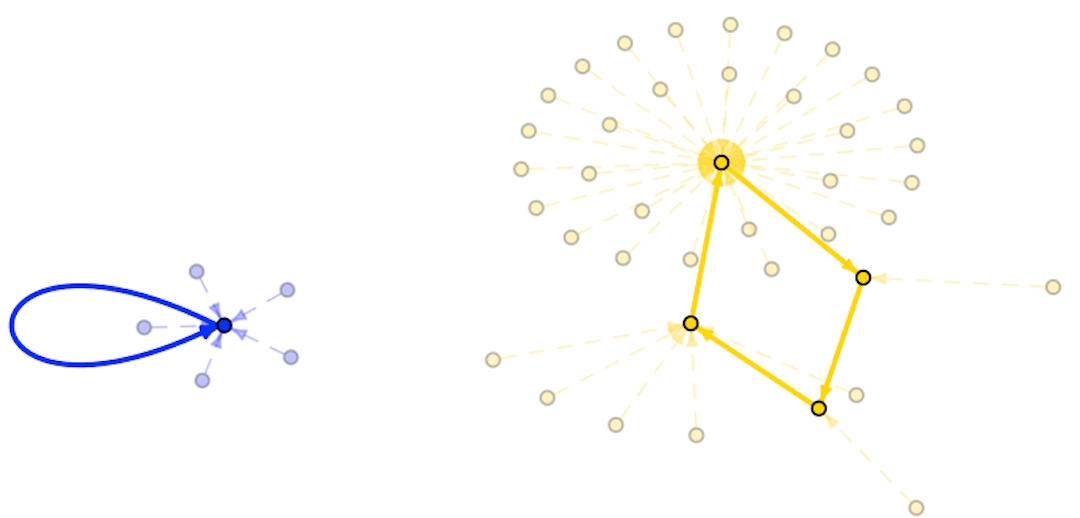}
\caption{The phase space graph of the \emph{tna} operon model with medium tryptophan and no glucose, $(\omega_e,\omega_{em},\gamma)=(0,1,0)$.}\label{fig:R_tna_med_noglu}
\end{figure}

Finally, let's turn to the parameter vector  $(\gamma,\omega_e,\omega_{em})=(0,0,1)$, of medium levels of external tryptophan in the absence of glucose. Unlike the other five cases of the parameter vector, the synchronous dynamics has more than one basin of attraction -- there are actually six. The smallest has just six nodes, leading into the fixed point $x^*\!=0010100$. There is another fixed point, $x^*\!=1111001$, which is contained in a basin with $10$ nodes. The $2$-cycle $1110000\longleftrightarrow 0011101$ is contained in a basin of $8$ states, and the last three basins all contain $4$-cycles. Specifically, the three $4$-cycles
\begin{align*}
& 0010000 \longrightarrow 0011100 \longrightarrow 1110100 \longrightarrow 0010101 \\
& 0011000 \longrightarrow 1111100 \longrightarrow 1110101 \longrightarrow 0010001 \\
& 1111000 \longrightarrow 1111101 \longrightarrow 1110000 \longrightarrow 0011001
\end{align*}
are contained in basins of $20$, $40$, and $44$ states, respectively. The phase space is shown in Figure~\ref{fig:R_tna_med_noglu}.

All four of these longer limit cycles are artifacts of synchrony, because they disappear in the asynchronous automaton, whose only two attractors are the fixed points described above. A medium range of tryptophan in the absense of glucose is where bistability might arise. 
Let's take a closer look into each of our fixed points, separately. The first one, $x^*\!=0010100$, represents the scenario when the cAMP-CAP protein is present, and there is a high likelihood of the Rho protein being bound to the \emph{tnaC} leader sequence. The operon should be off under these conditions due to the termination caused by the Rho protein. Indeed, all other variables are zero. For the second fixed point, $x^*\!=1111001$, all genes products are present, and so the operon is on. The variable $P=0$ is expected because the Rho protein should not terminate transcription, and $W=0$ is expected because tryptophan levels are only moderate, rather than high. In other words, under medium levels of tryptophan, it is possible for the \emph{tna} operon to be either induced or uninduced. This ``bistable range'' depends on both extracellular glucose and tryptophan levels, and the experimental results estimating this range is shown in \cite[Figure 9]{orozco2019bistable}. Of course, our Boolean model is not capable of quantifying this range; only in demonstrating that it is theoretically possible, using a much different framework than that of \cite{orozco2019bistable}. This gives further mathematical evidence that the \emph{tna} operon can exhibit bistability.

Table~\ref{tbl:tna-fixed-points} summarizes the fixed points reached from all six possible parameter vectors, and the corresponding state of the operon. In each case, the fixed point(s) reached is exactly what we expect, and furthermore, our model is able to predict the operon's bistability. 

\begin{table}[!ht]
\begin{tabular}{|c|c|c|}
\hline  Parameter vector & Fixed point(s) & Operon \\  $(\gamma,\omega_e,\omega_{em})$ & 
$\;\;(A,B,C,M,P,W,W_m)$ & state \\ \hline 
$(1,1,1)$ & $(0\,,\,0\,,\,0\,,\,0\,,\,0\,,\,0\,,\,1)$ & OFF \\ \hline 
$(1,0,0)$ & $(0\,,\,0\,,\,0\,,\,0\,,\,1\,,\,0\,,\,0)$ & OFF \\ \hline 
$(1,0,1)$ & $(0\,,\,0\,,\,0\,,\,0\,,\,1\,,\,0\,,\,0)$ & OFF \\ \hline
$(0,1,1)$ & $(1\,,\,1\,,\,1\,,\,1\,,\,0\,,\,1\,,\,1)$ & ON \\ \hline
$(0,0,0)$ & $(0\,,\,0\,,\,1\,,\,0\,,\,1\,,\,0\,,\,0)$ & OFF \\ \hline 
$(0,0,1)$ & $(0\,,\,0\,,\,1\,,\,0\,,\,1\,,\,0\,,\,0)$ & OFF \\ 
 & $(1\,,\,1\,,\,1\,,\,1\,,\,0\,,\,0\,,\,1)$ & ON \\ \hline
\end{tabular}
\caption{Fixed points of the \emph{tna} operon for each of the six parameter vectors.} \label{tbl:tna-fixed-points}
\end{table}

\section{A coupled model of the \emph{trp} and \emph{tna} operons}\label{sec:trp-tna}

\subsection{Model proposal and justification}

In the previous two sections, we modeled the \emph{trp} and \emph{tna} operons individually. In both cases, the fixed points match what we would expect biologically. This provides evidence that our models are capturing the basic biological functions, and corroborates the \emph{tna} operon's hypothesized bistability. That said, these operons do not exist independently of each other. For example, in our second model, if there is no glucose or extracellular tryptophan, then the \emph{tna} operon should be off. However, as tryptophan is an essential amino acid to all living organisms, any cell that does not have access to it will die. Rather than an \emph{E. coli} cell being stuck in the unique fixed point reached when $(\gamma,\omega_i,\omega_{im})=(0,0,0)$ and dying, its \emph{trp} operon should kick on and drive the system toward homeostasis. In other words, we should \emph{never} observe a fixed point with $W=W_m=0$ in a living cell. In this section, we will incorporate the key features of the \emph{trp} and \emph{tna} operons into a single Boolean model of the transport, synthesis, and metabolism of tryptophan in in \emph{E. coli}.

Our large model will have the same three parameters as the \emph{tna} model did: $\gamma$ denoting glucose, $\omega_e$ for high levels of extracellular tryptophan, and $\omega_{em}$ representing at least medium levels of extracellular tryptophan. Once again, we will assume that a glucose permease is always available. We will also assume in the absence of glucose, the other carbon sources preferred above tryptophan are collectively insufficient. In other words, if glucose is not available, then tryptophan should be metabolized. 

We will retain the same Roman letters for the variables as we did in the previous sections, except that we have to distinguish the different messenger RNAs. We will use $M_1$ to denote \emph{trp} mRNA and $M_2$ to denote \emph{tna} mRNA. Our thirteen variables are listed below.

\begin{multicols}{2}
\begin{enumerate}
    \item[$x_1=$] $A$: TnaA protein (tryptophanase)
    \item[$x_2=$] $B$: TnaB permease
    \item[$x_3=$] $C$: cAMP-CAP protein complex
    \item[$x_4=$] $E$: anthranilate synthase (high)
    \item[$x_5=$] $E_m$: anthranilate synthase (med)
    \item[$x_6=$] $L$: 3-4 loop (attenuation)
    \item[$x_7=$] $M_1$: \emph{trp} mRNA
    \item[$x_8=$] $M_2$: \emph{tna} mRNA
    \item[$x_9=$] $P$: Rho termination protein
    \item[$x_{10}=$] $R$: activated TrpR repressor protein
    \item[$x_{11}=$] $T$: charged tRNA
    \item[$x_{12}=$] $W$: intracellular tryptophan (high)
    \item[$x_{13}=$] $W_m$: intracellular tryptophan (med).
\end{enumerate}
\end{multicols}

One key difference in our large coupled model is the role of tryptophan: we have parameters for extracellular levels and variables for intracellular levels. Variables $x_1,\dots,x_8$ have local functions that do not involve tryptophan, and so we will retain the same functions in this model as we did for the \emph{trp} and \emph{tna} operons. However, for the remaining five variables, $x_9,\dots,x_{13}$, we need to propose new functions, which we will do below.

\begin{itemize}
    \item There is a near certainty of finding the Rho termination protein bound to the \emph{tnaC} leader sequence if intracellular tryptophan is not present. Therefore, its Boolean function is  $f_P=\Not{W}\And\Not{W_m}$.
    \item There will be sufficiently many activated repressor protein molecules in the cell to block transcription if there are high levels of tryptophan in the cell. The Boolean function is thus $f_R=W$. 
    \item Charged tRNA molecules will be available if there is some tryptophan present in the cell. The Boolean function is thus $f_T=W_m$.
    \item There are two ways that there can be high levels of tryptophan in the cell: (i) \emph{trp} mRNA is present and not attenuated, and there are not high levels of tryptophanse (TnaA) to metabolize it, (ii) there are high levels of extracellular tryptophan and the TnaB permease available to transport it. 
    \item There are four ways that there can be at least medium levels of intracellular tryptophan: (i) \emph{trp} mRNA is present, (ii) there are already high levels of intracellular tryptophan, which will not fully degrade or be utilized for other cellular functions right away, (iii) there is some extracellular tryptophan, and the TnaB permease is available to transport it, or (iv) there are high levels of extracellular tryptophan, some of which will get brought into the cell by the Mtr or AroP permeases. The Boolean function is thus
    $f_{W_m}=M_1\Or W\Or(B\And \omega_{em})\Or\omega_e$.
\end{itemize}

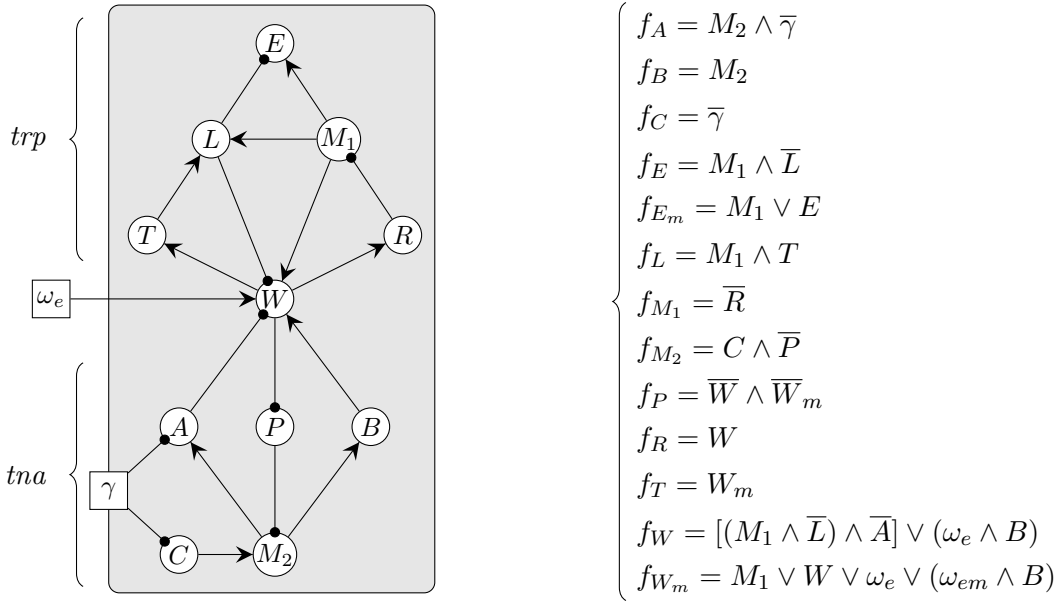
\begin{figure}[!ht]
  \tikzstyle{v} = [circle,fill=white,draw,inner sep=0pt, minimum size=5mm] 
  \tikzstyle{w} = [rectangle, fill=white,draw,inner sep=0pt, minimum size=5mm] 
  \tikzstyle{act} = [draw, -stealth]
  \begin{tikzpicture}[scale=1.7]
   \begin{scope}[shift={(0,0)}]
    \draw[fill=black!10, rounded corners] 
    (-1.3,-.3) rectangle (1.25,4.3);
    \node[v] (E) at (0,4) {\small $E$};
    \node[v] (L) at (-.5,3.25) {\small $L$};
    \node[v] (M1) at (.5,3.25) {\small $M_1$};
    \node[v] (T) at (-1,2.5) {\small $T$};
    \node[v] (R) at (1,2.5) {\small $R$};
    \node[v] (W) at (0,2) {\small $W$};
    \node[v] (A) at (-.75,1) {\small $A$};
    \node[v] (P) at (0,1) {\small $P$};
    \node[v] (B) at (.75,1) {\small $B$};
    \node[v] (C) at (-.75,0) {\small $C$};
    \node[v] (M2) at (0,0) {\small $M_2$};
    \node[w] (We) at (-1.75,2) {\small $\omega_e$};
    \node[w] (Ge) at (-1.3,.5) {\small $\gamma$};
    \draw[inhib] (R) to (M1); 
    \draw[activ] (M1) to (E);
    \draw[activ] (M1) to (L);
    \draw[activ] (M1) to (W);
    \draw[activ] (W) to (R);
    \draw[activ] (W) to (T);
    \draw[inhib] (L) to (E);
    \draw[inhib] (L) to (W);
    \draw[activ] (T) to (L);
    \draw[activ] (We) to (W); 
    \draw[inhib] (W) to (P);
    \draw[inhib] (A) to (W);
    \draw[activ] (B) to (W);
    \draw[activ] (C) to (M2);
    \draw[inhib] (P) to (M2);
    \draw[activ] (M2) to (A);
    \draw[activ] (M2) to (B);
    \draw[inhib] (Ge) to (A);
    \draw[inhib] (Ge) to (C);
    \draw[decorate,decoration={brace,amplitude=5pt}] 
      (-1.5,2.3) --  (-1.5,4.2); 
      \node[anchor=east] at (-1.7,3.25) {\emph{trp}};
     \draw[decorate,decoration={brace,amplitude=5pt}] 
      (-1.5,-.25) --  (-1.5,1.5); 
      \node[anchor=east] at (-1.7,.625) {\emph{tna}};
    \end{scope}
    \begin{scope}[scale=.9,shift={(3.05,2.2)}]
      \node[anchor=west] at (0,2.4) {$f_A=M_2\And\Not\gamma$};
      \node[anchor=west] at (0,2) {$f_B=M_2$};
      \node[anchor=west] at (0,1.6) {$f_C=\Not\gamma$};
      \node[anchor=west] at (0,1.2) {$f_E=M_1\And\Not{L}$};
      \node[anchor=west] at (0,.8) {$f_{E_m}=M_1\Or E$};
      \node[anchor=west] at (0,.4) {$f_L=M_1\And T$};
      \node[anchor=west] at (0,0) {$f_{M_1}=\Not{R}$};
	  \node[anchor=west] at (0,-.4) {$f_{M_2}=C\And\Not{P}$};
      \node[anchor=west] at (0,-.8) {$f_P=\Not{W}\And\Not{W_m}$};
      \node[anchor=west] at (0,-1.2) {$f_R=W$};
      \node[anchor=west] at (0,-1.6) {$f_T=W_m$};
      \node[anchor=west] at (0,-2) {$f_W=[(M_1\And\Not{L})\And\Not{A}]\Or (\omega_e\And B)$};
      \node[anchor=west] at (0,-2.4) {$f_{W_m}=M_1\Or W\Or \omega_e\Or(\omega_{em}\And B)$};
      \draw[decorate,decoration={brace,amplitude=5pt}]
      (0,-2.6) --  (0,2.6); 
    \end{scope}
  \end{tikzpicture}
  \caption{Our Boolean model of the \emph{tna} and \emph{trp} operons and its wiring diagram. Once again, the cells is shaded, variables are denoted by circles, parameters by squares, and pairs of variables representing high and medium levels have been collapsed into single nodes.}\label{fig:trp-tna-wiring-diagram}
\end{figure}

A wiring diagram of our coupled model is shown in Figure~\ref{fig:trp-tna-wiring-diagram}. As in the previous sections, the cell is shaded, variables are represented by circles and parameters by squares, and those representing multiple levels of the same molecular species are collapsed into a single node. Notice how the \emph{trp} and \emph{tna} operons are fairly separate, linked together by intracellular tryptophan in the middle, which one operon produces and the other consumes.

\subsection{Dynamics of the coupled model}

The phase space of our Boolean model has $2^{13}=8192$ nodes, which is still very manageable for BoolNet to compute, and even to qualitatively visualize. We need to consider six cases separately: either glucose is present or absent, and then extracellular tryptophan levels can be high, medium, or low. In every case, there is a unique fixed point, shown in Figure~\ref{tbl:big-fixed-points}. For three of these cases, there are also longer limit cycles under a synchronous update, but these disappear in the asynchronous automaton.  

\begin{table}[!ht]
\begin{tabular}{|c|c|c|c|}
\hline  Parameter vector & Fixed point & \emph{trp} & \emph{tna}\\ 
 $(\gamma,\omega_e,\omega_{em})$ & 
$\,\,\,(A,\,B,\,C,\,E,\,E_m,L,M_1,M_2,P,\,R,\,T,\,W,\,W_m)$ & operon & operon \\ \hline 
$(1,1,1)$ & $(0\;,\;0\;,\;0\;,\;0\;,\;1\;,\;1\;,\;1\;,\;0\;,\;0\;,\;0\;,\;1\;,\;0\;,\;1)$ & ON & OFF \\ 
$(1,0,0)$ & $(0\;,\;0\;,\;0\;,\;0\;,\;1\;,\;1\;,\;1\;,\;0\;,\;0\;,\;0\;,\;1\;,\;0\;,\;1)$ & ON & OFF \\ 
$(1,0,1)$ & $(0\;,\;0\;,\;0\;,\;0\;,\;1\;,\;1\;,\;1\;,\;0\;,\;0\;,\;0\;,\;1\;,\;0\;,\;1)$ & ON & OFF \\ \hline
$(0,1,1)$ & $(1\;,\;1\;,\;1\;,\;0\;,\;0\;,\;0\;,\;0\;,\;1\;,\;0\;,\;1\;,\;1\;,\;1\;,\;1)$ & OFF & ON \\ \hline 
$(0,0,0)$ & $(1\;,\;1\;,\;1\;,\;0\;,\;1\;,\;1\;,\;1\;,\;1\;,\;0\;,\;0\;,\;1\;,\;0\;,\;1)$ & ON & ON \\ 
$(0,0,1)$ & $(1\;,\;1\;,\;1\;,\;0\;,\;1\;,\;1\;,\;1\;,\;1\;,\;0\;,\;0\;,\;1\;,\;0\;,\;1)$ & ON & ON \\ \hline
\end{tabular}
\caption{The fixed points corresponding to the six parameter vectors in our coupled model.} \label{tbl:big-fixed-points}
\end{table}

One observation from Table~\ref{tbl:big-fixed-points} is that in all six cases, the \emph{trp} operon is on, but not at the highest levels, because $W=0$ and $W_m=1$. Another observation is that the fixed points are all the same if glucose is available. It turns out that these cases have the most interesting dynamics, and also exhibit artifacts of synchrony, which we will discuss below. 

\begin{figure}[!ht]
\includegraphics[width=1.8in]{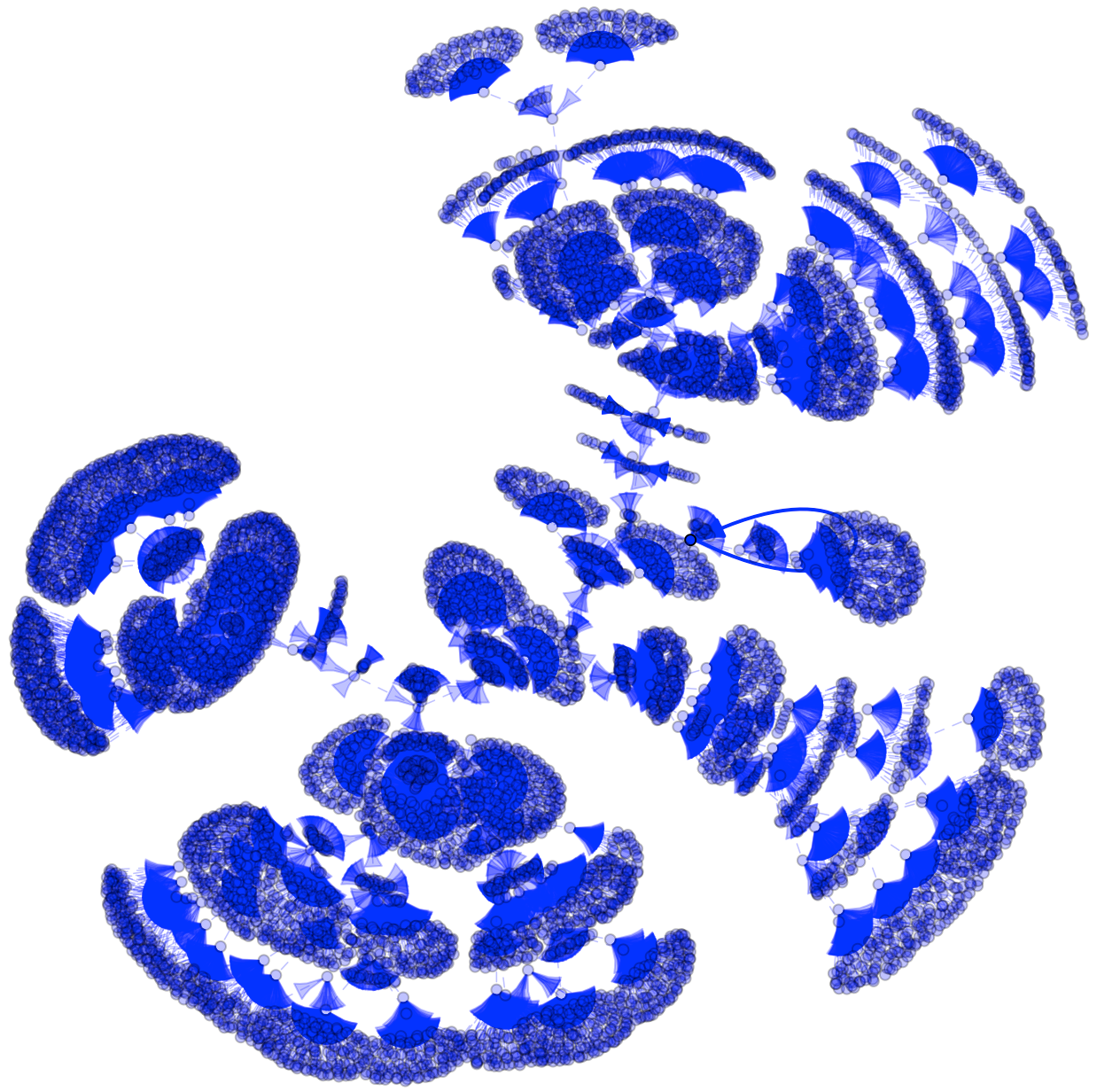}~\hspace{5mm}~\includegraphics[width=1.8in]{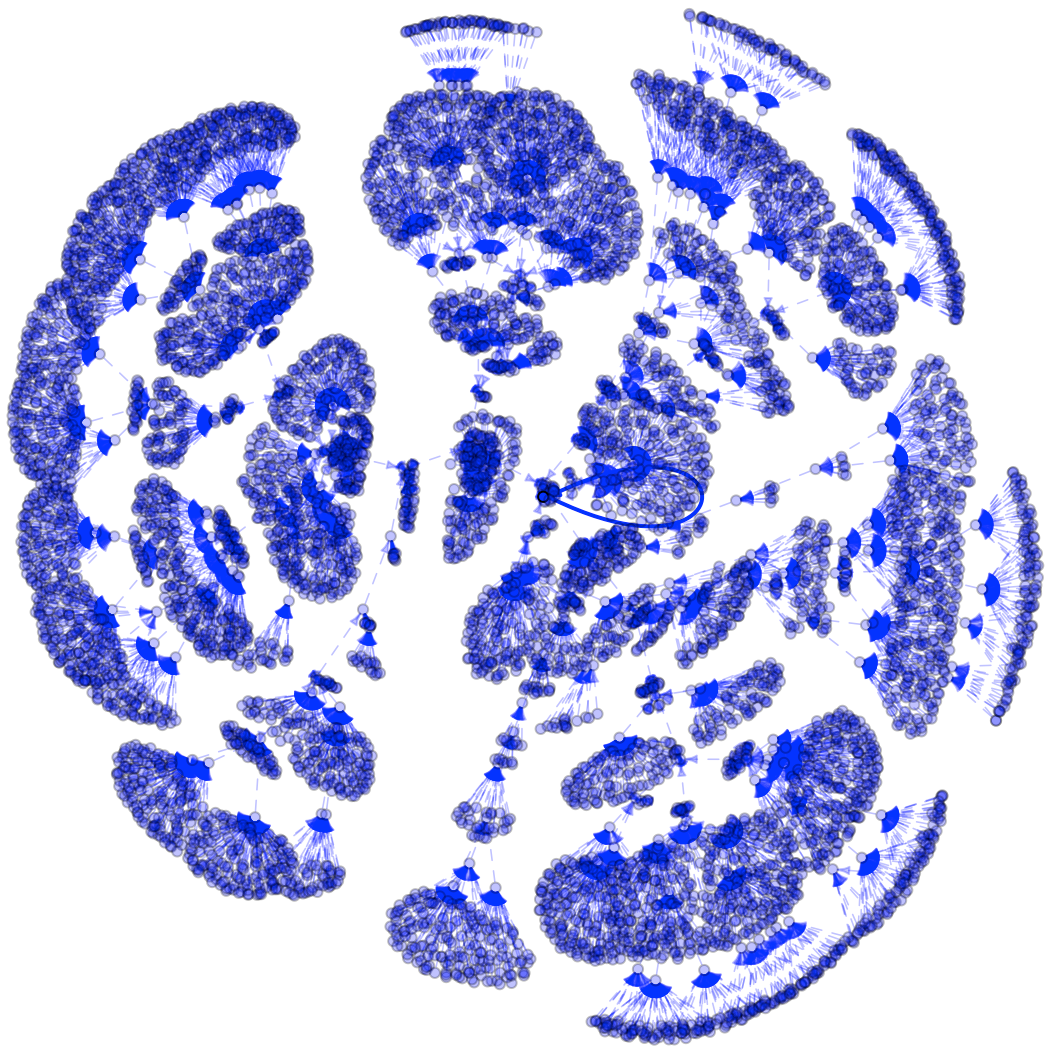}~\hspace{3mm}\includegraphics[width=1.8in]{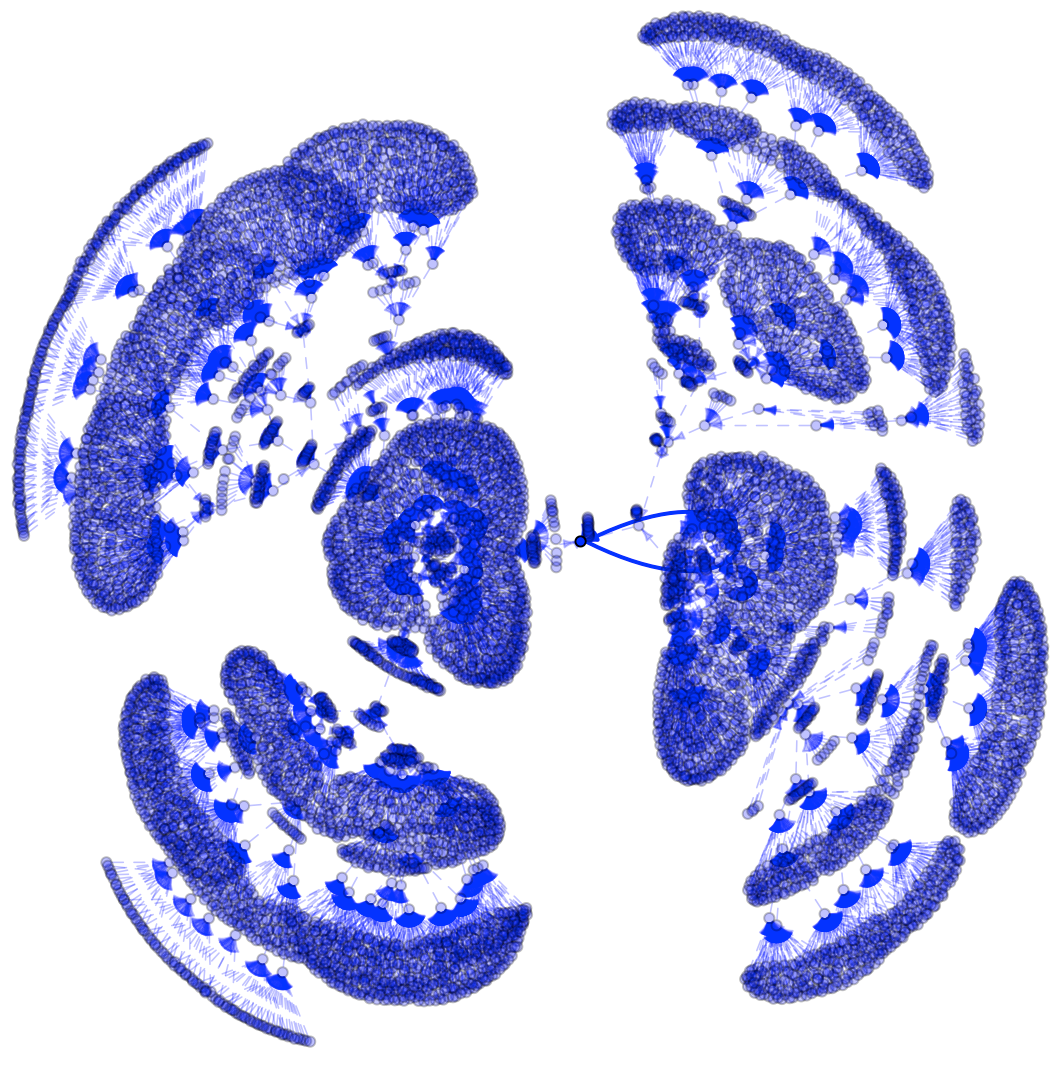}
\caption{Phase spaces of our \emph{tna} operon Boolean model in the absence of glucose: $\gamma=0$. From left to right are the cases where external tryptophan is low, medium and high. That is, $(\omega_e,\omega_{em})=(0,0)$, $(0,1)$, and $(1,1)$. In all three cases, the (blue) fixed point is what we should expect biologically.}\label{fig:R_big_noglu}
\end{figure}

Let's begin by analyzing the three parameter vectors that represent glucose being unavailable ($\gamma=0$). In all of these, the phase space consists of a single basin of attraction with the fixed point listed in Table~\ref{tbl:big-fixed-points}. These phase spaces are shown in Figure~\ref{fig:R_big_noglu}.

Now, we will explain the fixed points biologically. 
Tryptophan will have to be available for the cell to survive, whether it is brought into the cell via permeases or synthesized by the \emph{trp} operon. In either case, the \emph{tna} operon needs to turn itself on to utilize tryptophan as a carbon source. Indeed, for all three possibilities for $(\omega_e,\omega_{em})$, there is a unique fixed point with $A=B=C=M_2=1$. 

If there are high levels of extracellular tryptophan ($\omega_e=\omega_{em}=1$), then the TnaB, AroP and Mtr permeases will transport it into the cell, and the \emph{trp} operon will remain off, which agrees with $M_1=0$ and $E=E_m=0$ in the corresponding fixed point from Table~\ref{tbl:big-fixed-points}. The tryptophan will charge tRNA ($T=1$) and since the \emph{tna} operon will not metabolize tryptophan, high levels will be maintained inside the cell. 

If there is no tryptophan outside of the cell ($\omega_e=\omega_{em}=0$), then the \emph{trp} operon will need to synthesize it. The \emph{tna} operon will metabolize some of it, reducing its levels but not eliminating it ($W=0$, $W_m=1$), and still charging tRNA ($T=1$). The \emph{tna} operon remains on, which agrees with $A=B=C=M_2$ in the fixed point. In the long term, the \emph{trp} operon also remains on, and the tryptophan it produces will elevate the probability of the 3-4 hairpin loop forming on the mRNA strand, causing attenuation. Note that a value of $L=1$ does not distinguish between such a probability being 20\% or 98\%, so in this case we have to use our judgement to reason why it should be close to $1$. While this is a limitation of a coarse-grained Boolean model such as ours, it should not be seen as an impediment for the utility of one. 

Finally, the case when there are moderate levels of tryptophan outside of the cell ($\omega_e=0$, $\omega_{em}=1$) leads to the same fixed point. This is reasonable, because in either case, the \emph{trp} operon needs to synthesize some tryptophan, because the cell needs to utilize it as a carbon source, which it does with the \emph{tna} operon. In other words, the difference between $(\omega_e,\omega_{em})=(0,0)$ and $(0,1)$ does not affect the long-term behavior. 

\begin{figure}[!ht]
\includegraphics[width=1.8in]{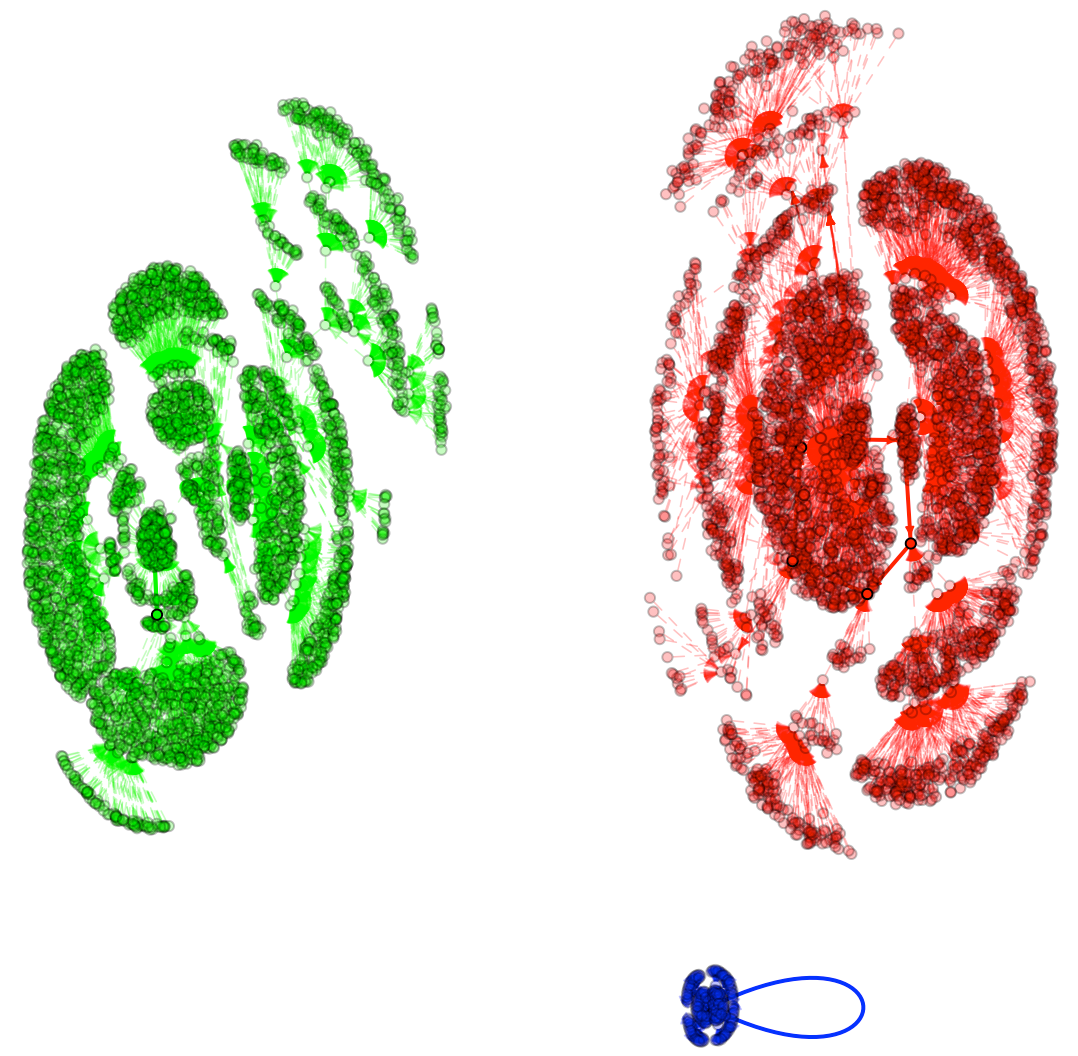}~\hspace{5mm}~\includegraphics[width=1.8in]{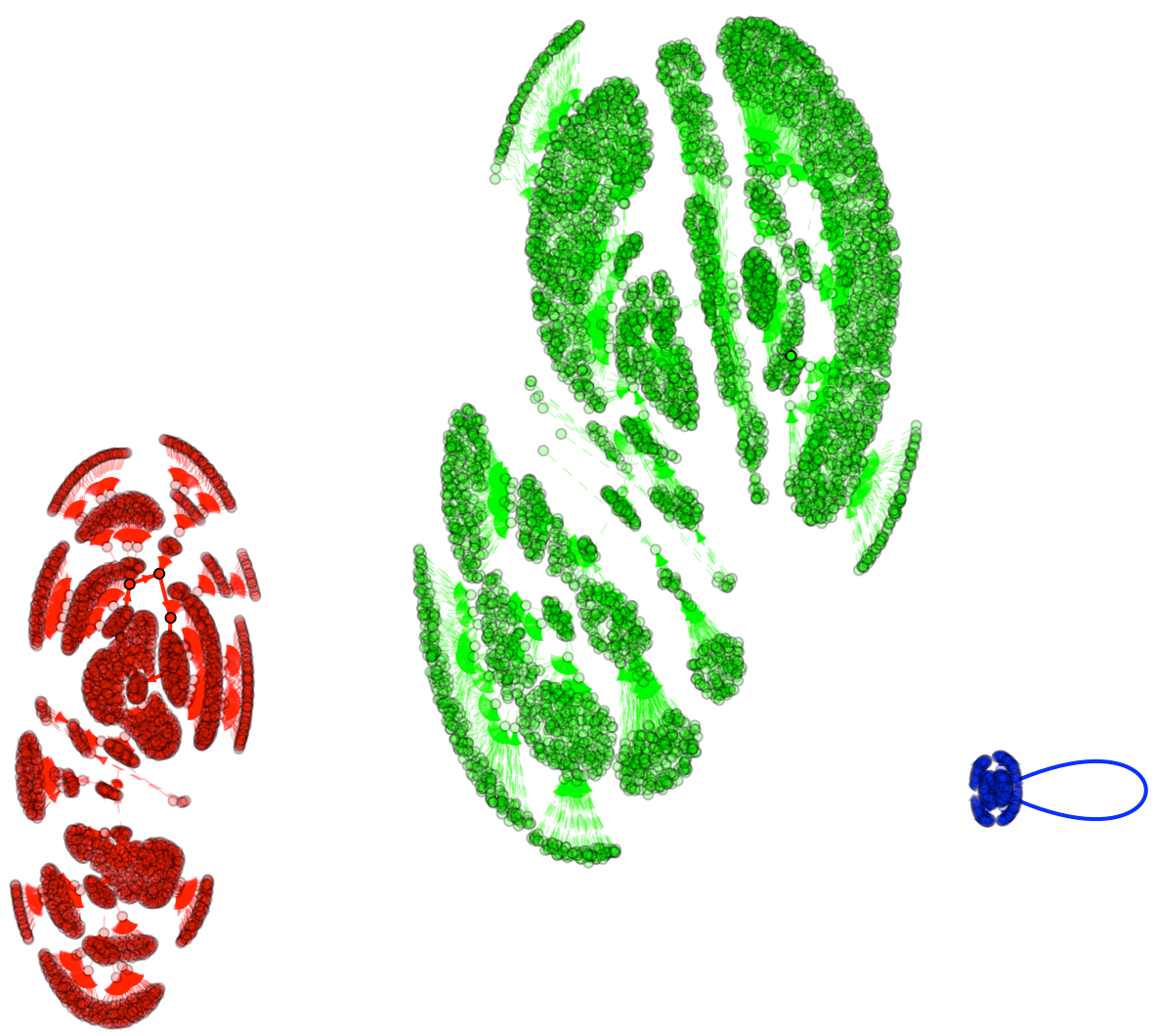}~\hspace{5mm}\includegraphics[width=1.9in]{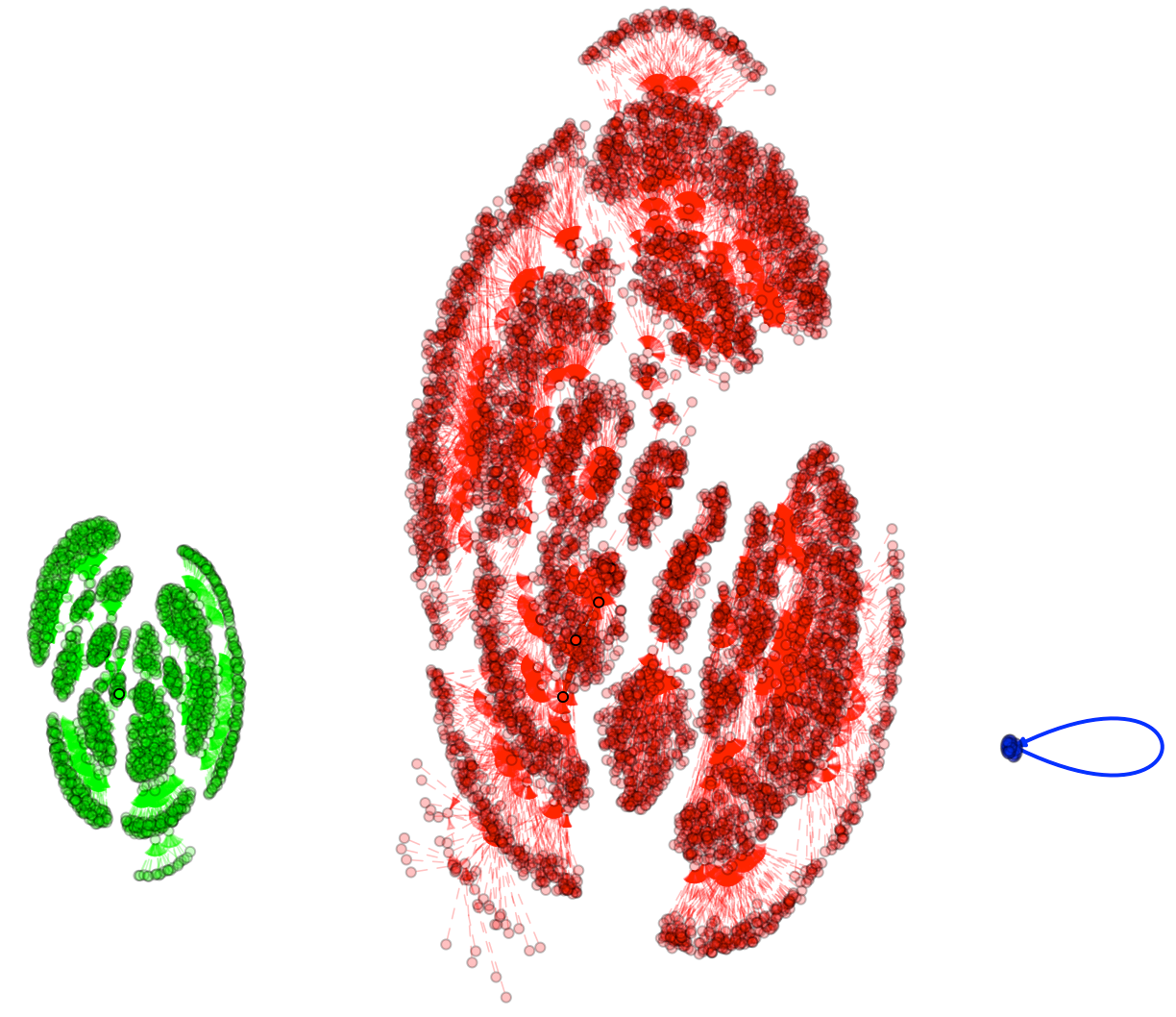}
\caption{Phase spaces of our coupled operon Boolean model with glucose: $\gamma=1$. From left to right are the cases where external tryptophan is low, medium and high. That is, $(\omega_e,\omega_{em})=(0,0)$, $(0,1)$, and $(1,1)$. In all three cases, the (blue) fixed point is what we should expect biologically.}\label{fig:R_big_glu}
\end{figure}

Now, we need to analyze the three parameter vectors that represent glucose being available ($\gamma=1$). In all of these, we expect the \emph{tna} operon should be off, since glucose will inhibit the cAMP-CAP protein complex. Indeed, we see that $A=B=M_2=0$ in the fixed point in Figure~\ref{tbl:big-fixed-points}. This means that the TnaB permease will not be translated, and so even if there are high levels of tryptophan outside of the cell, there will not be high levels inside of the cell. As a result, the \emph{trp} operon will be utilized to synthesize extra tryptophan. However, since it is not needed as a carbon source, it will not be produced at the highest levels. Excess tryptophan is more likely to charge tRNA than to bind to the repressor protein. As it result, it will elevate the probability of transcribed mRNA being attenuated, but not bind to sufficiently many repressor protein molecules to completely block transcription. This results in a fixed point of $M_1=L=1$ and $R=0$, and moderate levels of tryptophan: $W=0$ and $W_m=1$. The phase spaces for all three cases of glucose are shown in Figure~\ref{fig:R_big_glu}.

For all three cases shown in Figure~\ref{fig:R_big_glu}, the phase space consists of three connected components, and the smallest one has the same fixed point, described above and in Table~\ref{tbl:big-fixed-points}. When tryptophan is high, i.e., $(\gamma,\omega_e,\omega_{em})=(1,1,1)$, then the basin of attraction containing this fixed point has only $36$ states, or $0.44\%$ of the entire phase space. For higher levels of tryptophan, this basin has $192$ states. 

All three phase spaces also have a basin of attraction with the $2$-cycle
$0000101001101\longleftrightarrow 0001110000111$. If tryptophan is low, then this has size $3840$. For medium levels, it has $5344$ states, and for high levels, it has $1796$ states. Finally, when tryptophan levels are high, the third attractor is a $4$-cycle defined by
\[
0000110000101 \longrightarrow  0000001000101 \longrightarrow 
 0001111000111 \longrightarrow 0000111001101,
\]
sitting in a basin of size 6360. For lower levels of tryptophan, this attractor is the $6$-cycle
\begin{align*}
0000001000000 &\longrightarrow 0001101000011 \longrightarrow 0001101001111 \longrightarrow \\
 &\longrightarrow 0001110001111  \longrightarrow 0000100001101  \longrightarrow 0000000000100.
\end{align*}
This attractor is contained in a basin with $2656$ states under medium tryptophan levels, and $4160$ states under low levels. All of these longer attractors are artifacts of synchrony, as they disappear under an asynchronous update. That is, in all three cases, the asynchronous automaton has a unique strongly connected component, which is the fixed point shown in Table~\ref{tbl:big-fixed-points}. This is easily verified in the BoolNet package of R using the commands described in Section~\ref{subsec:trp-dynamics}, modified accordingly, such as replacing ``\verb+startStates=256+'' with ``\verb+startStates=8192+''.
 
We will finish this section with a remark about the curious observation that $P=0$ in all six fixed points, and why this does not contradict the fact that \emph{tna} mRNA is transcribed in some cases and not in others. There are two reasons for this. First, the value of $P=1$ describes the situation when the probability of the \emph{tnaC} leader sequence being bound to the Rho termination protein is near $1$, and this only happens is tryptophan is truly absent. Recall the analogy of a water sealant; $P=0$ can describe the probability being high but less than $1$ -- i.e., there is some tryptophan in the cell, which there should always be if we have functioning \emph{trp} and \emph{tna} operons to produce and/or transport it. If the \emph{tna} operon is on, as it is in the last three fixed points in Table~\ref{tbl:big-fixed-points}, then the probability of the repressor protein being bound to the \emph{tnaC} leader sequence should be low, and then is no problem here. In contrast, the \emph{tna} operon is off in the first three fixed points, despite $P=0$. However, glucose is present in all of these, and it inhibits the cAMP-CAP protein complex, which is required for transcription to initiate. In other words, even if the Rho protein is detached from the \emph{tnaC} leader sequence for some of the time, the \emph{tna} operon will remain off.

\section{Concluding remarks}

We will close this paper with a brief summary of this project, and some general questions about the ``artifacts of synchrony,'' such as characterizing when they occur, and detecting them algebraically. 

\subsection{Summary}

In this paper, we developed a Boolean model for the \emph{trp} operon in \emph{E. coli}, one of the most well-known in molecular biology, arguably only second to the \emph{lac} operon. As far as we know, our \emph{trp} operon model is the first Boolean model of a repressible operon. We also developed a Boolean model of the \emph{tna} operon, which was recently modeled with a system of ODEs that had two stable fixed points under a range of parameters for medium levels of tryptophan. Our model also supports the theory that this operon can exhibit bistability. Finally, we put our two models together to get a 13-variable  model of the synthesis, transport, and metabolism of tryptophan. This coupled model no longer had bistability, which can be explained by thinking of the \emph{trp} and \emph{tna} operons, which produce and consume tryptophan, respectively, as driving the system toward homeostasis. 

The models developed in this paper presented a unique set of challenges not found with similar Boolean models of the \emph{lac} and \emph{ara} operons. Capturing repression by both attenuation and a repressor protein was a challenge, and in order to do it, we had to carefully specify what the Boolean values of $0$ and $1$ represented, and they were different for both repression mechanisms. For attenuation, $L=0$ meant that the mRNA formed a 3-4 termination hairpin loop with an approximately zero probability, i.e., the ``floodgates were open.'' In contrast, $L=1$ represented a range of non-zero probabilities, and this simply cannot distinguish between a moderate probability that allows some full mRNA strands to be transcribed, but not at the highest expression level, vs. a very high probability that effectively shuts down the operon. A similar situation occurred in the \emph{trp} operon, where $P=1$ meant that the Rho termination protein was almost certainly bound to the \emph{tnaC} leader sequence, essentially shutting off transcription, whereas $P=0$ represented a wide range of situations of non-basal expression levels. 

One drawback of many mathematical models of molecular networks is that they view a system in isolation, and even if their results are supported by experimental data, such conditions are often contrived and would not exist in a living cell outside of the laboratory. For example, both mathematical modelling and experimental results have provided strong evidence that the \emph{lac} operon exhibits bistability when induced with artificial analogs of lactose. However, \cite{zander2017bistability} has argued recently why this operon might not be bistable when induced with the native sugar. The experimental and mathematical evidence for the bistability of the \emph{tna} operon does not preclude a similar situation or debate for this system.

\subsection{Detecting artifacts of synchrony}

The fixed points of a Boolean model are independent of whether the functions are updated synchronously or asynchronously. Both our \emph{tna} Boolean model, and the prior \emph{ara} operon model from \cite{jenkins2017bistability} have the curious property that under every parameter vector, the expected fixed point(s) are reached, but there are also are larger cycles under a synchronous update that disappear in the asynchronous automaton. We call these \emph{artifacts of synchrony}, and would like to pose several open-ended questions about why and when they appear. 

In theory, the asynchronous automaton could contain larger attractors that disappear under a synchronous update (the opposite of what we see), or they could both contain non-fixed point attractors. It is curious that this does not occur in the \emph{ara}, \emph{trp}, or \emph{tna} operons. There has been a great deal of work focused on understanding necessary and sufficient conditions for certain dynamical properties of Boolean networks. The most well-known of these includes a theorem of Remy, Ruet, and Thieffry that a positive feedback loop in the wiring diagram is necessary for multiple fixed points \cite{remy2008graphic}, a theorem of Richard that a negative feedback loop is a necessary condition for larger limit cycles \cite{richard2010negative}, and a classic result of Robert that if the wiring diagram is acyclic, then the phase space is \emph{nilpotent} -- it has a unique basin of attraction and fixed point \cite{robert1980iterations}.

Formally, we can say that a Boolean model $(f_1,\dots,f_n)$ has an \emph{artifact of synchrony} if it has a periodic cycle of length $k>1$ under a synchronous update, but only fixed points in its asynchronous automaton. We would like to pose the open-ended question about what are some necessary and sufficient conditions of the functions and/or signed wiring diagrams to observe dynamics with artifacts of synchrony.

Another open-ended question is how to detect such artifacts, or more generally, longer limit cycles, using techniques from computational algebra, which have been used to find fixed points. For networks of the size studied in this paper, this is not necessary, because they are small enough that BoolNet can simulate them in their entirely. However, there are many well-known larger Boolean models, such as 
a $23$-node model of the T-helper cell differentiation network \cite{saez-rodriguez2007logical}, or a $60$-node model of the segment polarity gene network in the \emph{Drosophila melanogaster} fruit fly \cite{albert2003topology}. Thus, it is both a practical question, as well as an interesting theoretical question, about how to develop algorithms using computational algebra to detect these properties. For example, it is relatively easy to detect fixed points algebraically. One can transform any Boolean function into a polynomial over $\F_2$, by replacing each $x\And y$ with $xy$, each $x\Or y$ with $x+y+xy$, and each $\Not{x}$ with $1+x$. Doing this yields a set of equations that we call an \emph{algebraic model}. An example of this for our \emph{tna} operon model is shown in Table~\ref{tbl:tna-polynomials}. 

\begin{table}[!ht]
\begin{tabular}{cclccccr}
    variable && local function $f_i$ &&&&& $f_i+x_i=0$ \\ \hline
    $A$ && $f_1=(1+\gamma)x_4$ &&&&& $x_1+(1+\gamma)x_4=0$ \\
    $B$ && $f_2=x_4$ &&&&& $x_2+x_4=0$ \\
    $C$ && $f_3=1+\gamma$ &&&&& $x_3+1+\gamma=0$ \\
    $M$ && $f_4=x_3(1+x_5)$ &&&&& $x_4+x_3(1+x_5)=0$  \\ 
    $P$ && $f_5=(1+x_6)(1+x_7)$ &&&&&
    $x_5+(1+x_6)(1+x_7)=0$ \\
    $W$ && $f_6=x_2\omega_e$ 
    &&&&& $x_2\omega_e+x_6=0 $ \\
    $W_m$ && $f_7=(x_6\omega_e + x_6 + \omega_e + 1)x_2\omega_{em}$ &&&&& $(x_6\omega_e + x_6 + \omega_e + 1)x_2\omega_{em}$\hspace{7mm} \\ 
    && $\hspace{7.5mm}+x_6+(1+x_6)\omega_e$ &&&&&  $+ x_6 + (1+x_6)\omega_e+x_7=0$ \\ \hline
\end{tabular}
\caption{The functions of our \emph{tna} model as polynomials, and the system of equations $\{f_i+x_i=0\mid i=1,\dots,7\}$ whose solutions are the fixed points.}\label{tbl:tna-polynomials}
\end{table}

The fixed points of an algebraic model are the solutions to the system of $\{f_i+x_i=0\mid i=1,\dots,n\}$ of nonlinear polynomials over $\F_2$. 
These polynomials for our \emph{tna} operon model are also shown in Table~\ref{tbl:tna-polynomials}, on the right. The solutions to such a system of nonlinear polynomials can be found easily by computing a Gr\"obner basis of the ideal in $\F_2[x_1,\dots,x_n]$ generated by these polynomials. 
This gives a simpler system of equations with the same solution. 
Using our \emph{tna} operon model above, this can be done by running the following commands in the computational algebra software package Macaulay2, which is freely available via the Macaulay2Web interface \cite{M2}, which we encourage the interested reader to try.
{\small
\begin{verbatim}
    R = ZZ/2[x1,x2,x3,x4,x5,x6,x7,x8];
    I = ideal(x1^2-x1,x2^2-x2,x3^2-x3,x4^2-x4,x5^2-x5,x6^2-x6,x7^2-x7,x8^2-x8); 
    Q = R/I;
    RingElement | RingElement :=(x,y)->x+y+x*y;
    RingElement & RingElement :=(x,y)->x*y;
    w_i = 0_Q;  w_im = 1_Q;
    f1 = (1+x3) & x4;  f2 = x1 | x4;  f3 = x4 & x6;  f4 = 1+x5;
    f5 = x7 | w_i;  f6 = x8 | w_im;  f7 = (1+x3) & x4;  f8 = x4 | x7;
    I = ideal(f1+x1, f2+x2, f3+x3, f4+x4, f5+x5, f6+x6, f7+x7, f8+x8);
    G = gens gb I
\end{verbatim}
}
The output of running the commands above is 
\begin{verbatim}
    (x3+1  x6  x7  x1+x7  x2+x7  x4+x7  x5+x7+1)
\end{verbatim}
which means that the Gr\"obner basis is
\begin{equation}\label{eqn:G}
   \mathcal{G}=\{x_3+1,\,x_6,\,x_1+x_7,\,x_2+x_7,\,x_4+x_7,\,x_5+x_7+1\}.
\end{equation}
Equating $x_i+x_j$ to zero over $\F_2$ simply means that $x_i=x_j$, whereas doing so for the term $x_i+x_j+1$ means that $x_i=x_j+1$. Therefore, the first two terms in Eq.~\eqref{eqn:G} tell us that $x_3=1$ and $x_6=0$, the next three tell us that $x_1=x_2=x_4=x_7$, and the last term tells us that $x_5$ is the negation of these values. In other words, the system $\{g=0\mid g\in\mathcal{G}\}$ has two solutions,
\[
x=(c,c,1,c,1+c,0,c),\qquad c\in\F_2.
\]
Notice that these are precisely the two fixed points in the bistable case, from Table~\ref{tbl:tna-fixed-points}. Computational algebra has been widely used in analyzing algebraic models \cite{dimitrova2007grobner, laubenbacher2004computational,laubenbacher2009computer}, and though it can easily identify the fixed points, the method described above would have to be modifed and extended to handle larger limit cycles, to eliminate their existence, or to identify when they are artifacts of synchrony. Exploring these questions is worthwhile not only for their potential utility to Boolean models, but also because they are
interesting in their own right, as a class of problems within computational algebra. The interested reader can find more details about this in \cite[Chapter 4]{robeva2018algebraic}, and about other new computational algebraic problems that have arisen from algebraic models of biological systems in \cite{macauley2020algebraic}.


\begin{thebibliography}{10}

\bibitem{albert2003topology}
R.~Albert and H.G. Othmer.
\newblock The topology of the regulatory interactions predicts the expression
  pattern of the segment polarity genes in {D}rosophila melanogaster.
\newblock {\em J. Theor. Biol.}, 223(1):1--18, 2003.

\bibitem{bliss1982role}
R.D. Bliss, P.R. Painter, and A.G. Marr.
\newblock Role of feedback inhibition in stabilizing the classical operon.
\newblock {\em J. Theor. Biol.}, 97(2):177--193, 1982.

\bibitem{dimitrova2007grobner}
E.S. Dimitrova, A.S. Jarrah, R.~Laubenbacher, and B.~Stigler.
\newblock A {G}r{\"o}bner fan method for biochemical network modeling.
\newblock In {\em Proc. Internat. Symposium Symb. Algebraic Comput.}, pages
  122--126. ACM, 2007.

\bibitem{gong2001mechanism}
F.~Gong, K.~Ito, Y.~Nakamura, and C.~Yanofsky.
\newblock The mechanism of tryptophan induction of tryptophanase operon
  expression: tryptophan inhibits release factor-mediated cleavage of
  {T}na{C}-peptidyl-t{RNAP}ro.
\newblock {\em Proc. Natl. Acad. Sci.}, 98(16):8997--9001, 2001.

\bibitem{goodwin1965oscillatory}
B.C. Goodwin.
\newblock Oscillatory behavior in enzymatic control processes.
\newblock {\em Adv. Enzym. Regul.}, 3:425--437, 1965.

\bibitem{M2}
D.R. Grayson and M.E. Stillman.
\newblock Macaulay2, a software system for research in algebraic geometry.
\newblock Available at \url{https://faculty.math.illinois.edu/Macaulay2/}.

\bibitem{gu2013knocking}
P.~Gu, F.~Yang, F.~Li, Q.~Liang, and Q.~Qi.
\newblock Knocking out analysis of tryptophan permeases in {E}scherichia coli
  for improving {L}-tryptophan production.
\newblock {\em Appl. Microbiol. Biotechnol.}, 97(15):6677--6683, 2013.

\bibitem{jacob1960l'operon}
F.~Jacob, D.~Perrin, C.~S{\'a}nchez, and J.~Monod.
\newblock L'op{\'e}ron: groupe de g{\`e}nes {\`a} expression coordonn{\'e}e par
  un op{\'e}rateur.
\newblock {\em C.R. Acad. Sci.}, 250:1727--1729, 1960.

\bibitem{jenkins2017bistability}
A.~Jenkins and M.~Macauley.
\newblock Bistability and asynchrony in a {B}oolean model of the {\sc
  l}-arabinose operon in escherichia coli.
\newblock {\em Bull. Math Biol.}, 79(8):1778--1795, 2017.

\bibitem{laubenbacher2004computational}
R.~Laubenbacher and B.~Stigler.
\newblock A computational algebra approach to the reverse engineering of gene
  regulatory networks.
\newblock {\em J. Theor. Biol.}, 229(4):523--537, 2004.

\bibitem{laubenbacher2009computer}
R.~Laubenbacher and B.~Sturmfels.
\newblock Computer algebra in systems biology.
\newblock {\em Amer. Math. Monthly}, pages 882--891, 2009.

\bibitem{li2014camp-independent}
G.~Li and K.D. Young.
\newblock A c{AMP}-independent carbohydrate-driven mechanism inhibits tna{A}
  expression and {T}na{A} enzyme activity in {E}scherichia coli.
\newblock {\em Microbiology}, 160(9):2079--2088, 2014.

\bibitem{li2015new}
G.~Li and K.D. Young.
\newblock A new suite of tna{A} mutants suggests that {E}scherichia coli
  tryptophanase is regulated by intracellular sequestration and by occlusion of
  its active site.
\newblock {\em BMC Microbiol.}, 15(1):1--17, 2015.

\bibitem{macauley2020algebraic}
M.~Macauley and R.~Robeva.
\newblock Algebraic models, inverse problems, and pseudomonomials from biology.
\newblock {\em Lett. Biomath.}, 7(1):81--104, 2020.

\bibitem{mackey2004modeling}
M.C. Mackey, M.~Santill{\'a}n, and N.~Yildirim.
\newblock Modeling operon dynamics: the tryptophan and lactose operons as
  paradigms.
\newblock {\em C. R. Biol.}, 327(3):211--224, 2004.

\bibitem{matsushiro1965transcription}
A.~Matsushiro, K.~Sato, J.~Ito, S.~Kida, and F.~Imamoto.
\newblock On the transcription of the tryptophan operon in {E}scherichia coli:
  {I}. the tryptophan operator.
\newblock {\em J. Mol. Biol.}, 11(1):54--63, 1965.

\bibitem{mussel2010boolnet}
C.~M{\"u}ssel, M.~Hopfensitz, and H.A. Kestler.
\newblock Bool{N}et--an {R} package for generation, reconstruction and analysis
  of {B}oolean networks.
\newblock {\em Bioinformatics}, 26(10):1378--1380, 2010.

\bibitem{nelson2005principles}
D.L. Nelson, M.M. Cox, and A.~L. Lehninger.
\newblock Principles of biochemistry.
\newblock {\em WH Freeman and Company, New York, fourth edition edition},
  1(1.1):2, 2005.

\bibitem{orozco2019bistable}
D.I. Orozco-G{\'o}mez, J.E. Sosa-Hern{\'a}ndez, {\'O}.A. Gallardo-Navarro,
  J.~Santana-Solano, and M.~Santill{\'a}n.
\newblock Bistable behaviour and medium-dependent post-translational regulation
  of the tryptophanase operon regulatory pathway in {E}cherichia coli.
\newblock {\em Sci. Rep.}, 9(1):5451, 2019.

\bibitem{remy2008graphic}
{\'E}.~Remy, P.~Ruet, and D.~Thieffry.
\newblock Graphic requirements for multistability and attractive cycles in a
  {B}oolean dynamical framework.
\newblock {\em Adv. Appl. Math.}, 41(3):335--350, 2008.

\bibitem{richard2010negative}
A.~Richard.
\newblock Negative circuits and sustained oscillations in asynchronous automata
  networks.
\newblock {\em Adv. Appl. Math.}, 44(4):378--392, 2010.

\bibitem{robert1980iterations}
R.~Robert.
\newblock Iterations sur des ensembles finis et automates cellulaires
  contractants.
\newblock {\em Linear Algebra Appl.}, 29:393--412, 1980.

\bibitem{robeva2013mathematical}
R.~Robeva and T.~Hodge.
\newblock {\em Mathematical concepts and methods in modern biology: using
  modern discrete models}.
\newblock Academic Press, 2013.

\bibitem{robeva2018algebraic}
R.~Robeva and M.~Macauley.
\newblock {\em Algebraic and Combinatorial Computational Biology}.
\newblock Elsevier, 2018.

\bibitem{saez-rodriguez2007logical}
J.~Saez-Rodriguez et~al.
\newblock A logical model provides insights into {T} cell receptor signaling.
\newblock {\em PLoS Comput. Biol.}, 3(8):e163, 2007.

\bibitem{santillan2001dynamic}
M.~Santill{\'a}n and M.C. Mackey.
\newblock Dynamic regulation of the tryptophan operon: a modeling study and
  comparison with experimental data.
\newblock {\em Proc. Natl. Acad. Sci.}, 98(4):1364--1369, 2001.

\bibitem{simao2005qualitative}
E.~Simao, E.~Remy, D.~Thieffry, and C.~Chaouiya.
\newblock Qualitative modelling of regulated metabolic pathways: application to
  the tryptophan biosynthesis in {E}. coli.
\newblock {\em Bioinformatics}, 21(suppl 2):ii190--ii196, 2005.

\bibitem{thomas1990biological}
R.~Thomas and R.~d'Ari.
\newblock {\em Biological feedback}.
\newblock CRC press, 1990.

\bibitem{veliz-cuba2011boolean}
A.~Veliz-Cuba and B.~Stigler.
\newblock Boolean models can explain bistability in the lac operon.
\newblock {\em J. Comp. Biol.}, 18(6):783--794, 2011.

\bibitem{yanofsky1991physiological}
C.~Yanofsky, V.~Horn, and P.~Gollnick.
\newblock Physiological studies of tryptophan transport and tryptophanase
  operon induction in {E}scherichia coli.
\newblock {\em J. Bacteriol.}, 173(19):6009--6017, 1991.

\bibitem{yanofsky1984repression}
C.~Yanofsky, R.L. Kelley, and V.~Horn.
\newblock Repression is relieved before attenuation in the trp operon of
  {E}scherichia coli as tryptophan starvation becomes increasingly severe.
\newblock {\em J. Bacteriol.}, 158(3):1018--1024, 1984.

\bibitem{yildirim2012mathematical}
N.~Yildirim.
\newblock Mathematical modeling of the low and high affinity arabinose
  transport systems in {E}scherichia coli.
\newblock {\em Mol. BioSyst.}, 8(4):1319--1324, 2012.

\bibitem{yildirim2004dynamics}
N.~Yildirim, M.~Santillan, D.~Horike, and M.C. Mackey.
\newblock Dynamics and bistability in a reduced model of the lac operon.
\newblock {\em Chaos}, 14(2):279--292, 2004.

\bibitem{zander2017bistability}
D.~Zander, D.~Samaga, R.~Straube, and K.~Bettenbrock.
\newblock Bistability and nonmonotonic induction of the lac operon in the
  natural lactose uptake system.
\newblock {\em Biophys. J.}, 112(9):1984--1996, 2017.

\bibitem{zheng2004identification}
D.~Zheng, C.~Constantinidou, J.L. Hobman, and S.D. Minchin.
\newblock Identification of the {CRP} regulon using in vitro and in vivo
  transcriptional profiling.
\newblock {\em Nucleic Acids Res.}, 32(19):5874--5893, 2004.

\end{thebibliography}

\end{document}